\documentclass[aps,twocolumn,superscriptaddress,nolongbibliography]{revtex4-2}
\usepackage{multirow}
\usepackage{graphicx}
\usepackage[utf8]{inputenc}
\usepackage[T1]{fontenc}
\usepackage{epstopdf}
\usepackage{textcomp}
\usepackage[usenames,dvipsnames]{color}
\usepackage{amsbsy}
\usepackage{amsmath}
\usepackage{amssymb}
\usepackage{amsfonts}
\usepackage{dsfont}
\usepackage{mathtools}
\usepackage{braket}
\usepackage{array}
\usepackage{gensymb}
\usepackage{placeins}
\usepackage{dashrule}
\usepackage{xcolor}
\usepackage{booktabs}
\usepackage{natbib}
\usepackage{mfirstuc}
\usepackage{comment}
\usepackage{bm}             
\usepackage[colorlinks=true,linkcolor=blue ,citecolor=blue,urlcolor=blue]{hyperref}
\setcitestyle{super}

\usepackage[normalem]{ulem}

\newcommand{\ten}[1]{\bm{#1}}           
\newcommand{\vco}[1]{\hat{\mathbf{#1}}} 

\newcommand{\vc}[1]{\boldsymbol{\mathrm #1}}

\newcommand{\matr}[1]{\ensuremath{\underline{\mathbf{#1}}}}

\newcommand{\AMATIS}{\texttt{AMATIS}}
\newcommand{\expo}[1]{\ensuremath{\mathrm{e}^{#1}}}

\makeatletter
\newcommand\sh@ftsym[1]{%
  \smash{\raise-.3ex\hbox{$\scriptscriptstyle#1$}}}
\newcommand\Gtrless{\mathbin{\sh@ftsym({\gtrless}\sh@ftsym)}}
\makeatother
\usepackage[singlelinecheck=false, justification=RaggedRight, format=plain]{caption}
\DeclareCaptionLabelSeparator{bar}{ $\vc{\mid}$ }
\renewcommand{\onlinecite}[1]{Ref.~\nocite{#1}\citenum{#1}} 
\makeatletter
\renewcommand{\section}{%
  \@startsection{section}{1}{0pt}%
    {2.0ex plus 1ex minus 0.2ex} 
    {1.0ex plus 0.2ex}           
    {\normalfont\bfseries\flushleft}} 
\renewcommand{\subsection}{%
  \@startsection{subsection}{2}{0pt}%
    {1.5ex plus .8ex minus .2ex}
    {0.8ex plus .2ex}
    {\normalfont\bfseries\flushleft}%
}
\renewcommand{\subsubsection}{%
  \@startsection{subsubsection}{3}{0pt}%
    {1.2ex plus .6ex minus .2ex}%
    {0.6ex plus .2ex}%
    {\normalfont\itshape\flushleft}%
}
\makeatother

\makeatletter
\def\my@tag@font{\normalsize}
\def\maketag@@@#1{\hbox{\m@th\normalfont\my@tag@font#1}}
\let\amsmath@eqref\eqref
\renewcommand{\eqref}[1]{%
  {\let\my@tag@font\relax\amsmath@eqref{#1}}%
}
\makeatother

 \allowdisplaybreaks
 
\begin{document}
 \title{Unbiased first-principles construction of complete tensorial spin Hamiltonians}

\newcommand{\miit}{Fert Beijing Institute, MIIT Key Laboratory of  Spintronics, School of Integrated Circuit Science and Engineering, Beihang University, Beijing, 100191, China}
\newcommand{\skl}{State Key Laboratory of Spintronics, Hangzhou International Innovation Institute, Beihang University, Hangzhou 311115, China}
\newcommand{\ec}{Engineering Department, Cambridge University, Cambridge CB2 1PZ, UK}
\newcommand{\fz}{Peter Gr\"unberg Institut, Forschungszentrum J\"ulich and JARA, 52425 J\"ulich, Germany}
\newcommand{\rwth}{Institute for Theoretical Physics,  RWTH Aachen University, 52074 Aachen, Germany}

\author{Haichang Lu}
\email{HaichangLu@buaa.edu.cn}
\affiliation{\miit}
\affiliation{\skl}
\affiliation{\ec}
\author{Boyang Deng}
\affiliation{\miit}
\affiliation{\skl}
\author{Hiroshi Katsumoto}
\affiliation{\fz}
\author{John Robertson}
\affiliation{\ec}
\author{Weisheng Zhao}
\affiliation{\miit}
\affiliation{\skl}
\author{Stefan Bl\"ugel}
\affiliation{\fz} 
\affiliation{\rwth}

\date{\today}

\begin{abstract}
Magnetic ground states are commonly predicted using spin Hamiltonians whose interaction terms are selected \emph{a priori}, potentially overlooking the microscopic interactions that govern complex magnetic order. 
Here, we introduce a general framework for the unbiased first-principles construction of symmetry-complete tensorial spin Hamiltonians and its automated implementation in \AMATIS. The framework constructs the
Hamiltonian directly from density-functional theory while rigorously enforcing quantum spin algebra and crystallographic symmetry. Applied to representative two-dimensional van der Waals magnets, the framework reproduces established magnetic interactions and uncovers hidden physics beyond conventional spin models, including chiral interactions that stabilize metastable skyrmions,  higher-rank tensorial interactions that reconstruct the magnetic phase diagram and establish stabilizing competing multi-$Q$ phases, and an emergent $p$-wave altermagnetic electronic structure. Our results demonstrate that unbiased tensorial Hamiltonian construction
provides a predictive alternative to the conventional practice of manually selecting spin-model interactions, enabling first-principles discovery of unconventional magnetic phases. 
\end{abstract}

\maketitle

\section*{Introduction}

Magnetism originates from the collective behavior of interacting quantum spins and underpins a broad range of phenomena in condensed-matter physics, including long-range magnetic order, spin excitations, topological magnetic textures, electronic topology, and magnetotransport. These phenomena form the basis of modern technologies ranging from spintronic devices and magnetic memories to quantum information platforms and energy-efficient sensors. Across these diverse systems, effective spin Hamiltonians provide the fundamental framework for describing magnetic interactions and predicting magnetic ground states and excitations. As magnetic materials become increasingly complex through reduced dimensionality, strong spin-orbit coupling, geometric frustration, and interface engineering, constructing predictive spin Hamiltonians has become one of the central challenges of modern condensed-matter physics.

Although first-principles electronic-structure methods based on density functional theory (DFT) accurately determine the electronic origin of magnetic interactions, translating this information into predictive spin Hamiltonians remains a major bottleneck. In practice, effective Hamiltonians are rarely derived in an unbiased manner. Instead, they are typically constructed by selecting \emph{a priori} a limited set of interaction terms—most commonly Heisenberg exchange, Dzyaloshinskii-Moriya interactions (DMI), and magnetic anisotropy—and subsequently fitting their parameters to first-principles calculations. While this strategy has proved successful for many materials, it inevitably assumes which microscopic interactions are relevant before the Hamiltonian is established. As magnetic systems become increasingly complex, with competing relativistic and higher-order interactions of comparable energy scales, such assumptions may bias the resulting spin model, obscure the microscopic origin of emergent magnetic phases, and ultimately limit predictive modelling.

The Heisenberg model, introduced nearly a century ago, remains the cornerstone of theoretical magnetism because its bilinear isotropic exchange interaction successfully describes the magnetic order and excitations of many collinear and non-collinear magnets.\cite{Heisenberg1928,VanVleck1937,Anderson1950,MoriyaYosida1953} However, numerous quantum materials—particularly low-dimensional, frustrated, and spin-orbit-coupled systems—are governed by a much richer hierarchy of interactions. These include higher-order multi-spin terms such as biquadratic,\cite{Kittel1960,Thouless1965} ring,\cite{Takahashi1977} and four-spin three-site exchange,\cite{Hoffmann2020} topological chiral-chiral interactions,\cite{Grytsiuk2020} anisotropic exchange including Dzyaloshinskii-Moriya interactions,\cite{Moriya1960,Dzyaloshinskii1957} chiral biquadratic interactions,\cite{Brinker2019,Laszloffy2019} and Kitaev or compass-type anisotropies,\cite{Jackeli2009,Nussinov2015} which may compete on similar energy scales and qualitatively reshape magnetic phase diagrams. Restricting spin Hamiltonians to a predefined subset of interactions can therefore conceal essential microscopic mechanisms responsible for unconventional magnetic order.

These developments have established tensorial spin Hamiltonians as the natural framework for describing magnetic interactions beyond conventional model Hamiltonians. Rather than representing magnetic couplings by isolated scalar or vector parameters, tensorial formulations express isotropic, antisymmetric, anisotropic, and higher-order interactions within a unified symmetry-constrained description. Recent advances based on Hubbard-type models,\cite{Hoffmann2020} perturbation theory, Green-function techniques,\cite{Brinker2019,Grytsiuk2020,Lounis2020} spin-cluster expansions,\cite{Drautz2004,Drautz2005,Szunyogh:2011,Hatanaka:2025} and cluster multipole approaches\cite{Bouaziz2025} have revealed an increasingly rich hierarchy of tensorial magnetic interactions, highlighting the need for general first-principles methods capable of identifying them systematically.

Several first-principles techniques have been developed to extract magnetic interactions from electronic-structure calculations, including linear-response methods,\cite{Liechtenstein1987,Bruno2003,MartinezCarracedo2023,Szilva2023} frozen-magnon approaches,\cite{Kurz2004} the disordered-local-moment method combined with spin-cluster expansions,\cite{Pindor1983,Szunyogh:2011,Bouaziz2025} and total-energy mapping schemes based on selected magnetic configurations.\cite{Grytsiuk2020,Xiang2013,Yang2015,Torelli2019,Sabani2020} Despite their considerable success, most existing approaches still require the form of the spin Hamiltonian to be specified before its parameters are determined. Consequently, the predictive capability of the resulting model remains intrinsically linked to prior assumptions about which interactions are included. A general first-principles framework that constructs symmetry-complete tensorial spin Hamiltonians directly from electronic structure remains unavailable.

To overcome this limitation, we develop a general framework for the unbiased first-principles construction of tensorial spin Hamiltonians. The framework systematically enumerates all symmetry-allowed non-relativistic and relativistic spin interactions up to a precision prescribed tensor rank and interaction range, and determines the independent tensor components consistent with crystallographic symmetry and quantum-mechanical spin algebra. Their interaction coefficients are extracted from reference-state-independent relativistic non-collinear DFT calculations using a constrained fitting procedure. We implement this framework in \AMATIS\ (Automated Magnetism Analysis via Tensorial Interacting Spins), providing an automated realization of the complete tensorial construction and its decomposition into isotropic, anisotropic, and higher-order magnetic interactions suitable for atomistic spin simulations and finite-temperature modelling.

Two-dimensional (2D) magnetic van der Waals materials provide particularly stringent tests for such an unbiased construction because reduced coordination, broken inversion symmetry, strong spin-orbit coupling, and competing magnetic interactions frequently occur on comparable energy scales. Consequently, simplified spin Hamiltonians often become insufficient to describe their magnetic energy landscape. We therefore employ chromium triiodide and chromium dichalcogenide monolayers as representative model systems to demonstrate how complete tensorial Hamiltonians reconstruct the hierarchy of microscopic magnetic interactions.

Application of the framework to representative two-dimensional magnetic systems reveals long-range isotropic and anisotropic rank-2 and rank-4 interactions whose competition stabilizes, depending on strain, a remarkably rich magnetic energy landscape comprising single-$Q$, double-$Q$, ferrimagnetic, topologically non-trivial sextuple-$Q$, and emergent $p$-wave altermagnetic states, together with isolated skyrmionic textures in van der Waals heterostructures. These results demonstrate that unbiased tensorial spin-Hamiltonian construction can fundamentally change the predicted magnetic phase diagram and provide a new paradigm for predictive first-principles modelling of complex magnetic materials. Thus, conventional magnetic interactions are therefore not assumed as the starting point of the construction, but emerge as physically interpretable projections of a symmetry-adapted tensorial representation derived from first-principles energies.

\section*{Results}
\subsection{A general framework for unbiased tensorial spin Hamiltonians}
\label{sec:tensorial_hamiltonian}

The predictive description of magnetic materials from first principles requires a spin Hamiltonian that does not assume the relevant magnetic interactions \emph{a priori}. While the Heisenberg model 
\begin{equation}
\hat{H}_{\mathrm{H}} = \sum_{i<j} J^{ij}\,\vco{s}_{i}\cdot\vco{s}_{j},
\end{equation}
has provided the foundation for the microscopic understanding of magnetism for nearly a century, its restriction to isotropic pairwise ($i, j$)  bilinear exchange between spin operators  $\vco{s}$ with scalar exchange constant $J^{ij}$  limits its applicability whenever competing exchange mechanisms, relativistic interactions, or higher-order spin couplings become comparable in magnitude. To overcome these limitations, we formulate a general tensorial spin Hamiltonian that systematically incorporates all symmetry-allowed magnetic interactions within a unified and unbiased framework. Conventional spin Hamiltonians consequently emerge as successive low-order approximations of this general expansion rather than being imposed from the outset. 

The microscopic description of interacting localized magnetic moments is conventionally based on the extended quantum Heisenberg Hamiltonian
\begin{equation}
\hat{H}=\sum_{i,j} \vco{s}_{i}^{\mathrm{T}} \cdot \ten{J}_{ij} \cdot \vco{s}_{j}=\sum_{i,j}\ten{J}_{(2)}^{i,j} : \vco{s}_{i}\otimes\vco{s}_{j}\, ,
\end{equation}
where $\vco{s}_{i}$ denotes the quantum spin operator at lattice site $i$, $\ten{J}_{ij}=\ten{J}_{(2)}^{i,j}\in\mathds{R}^3\otimes \mathds{R}^3$ is the generalized exchange rank-2 Cartesian tensor acting in the three-dimensional spin space and coupling spins located at sites $i$ and $j$, ``:'' denotes double contraction in spin space, and the summation over $i,j$ excludes duplicated interactions while including the case $i=j$. In its most general form, $\ten{J}_{ij}$ contains the isotropic Heisenberg exchange, the antisymmetric Dzyaloshinskii--Moriya interaction, and the symmetric anisotropic exchange interaction. Despite its broad success, this bilinear description neglects higher-order magnetic interactions that become increasingly important in frustrated magnets, itinerant systems, and materials exhibiting complex non-collinear or multi-$Q$ magnetic order.

To establish a complete microscopic description without introducing assumptions regarding the form of the magnetic interactions, we expand the magnetic energy in terms of tensor products of spin operators. The resulting quantum Hamiltonian is written as
\begin{equation}
\hat{H} = \sum_{n=1}^{\infty} \hat{H}^{(2n)}, 
\end{equation}
where each rank-$2n$ contribution is given by
\begin{equation}\label{eq:tensorial_spin_hamiltonian}
\hat{H}^{(2n)} = \sum_{i_1, \ldots, i_{2n}} \ten{J}_{(2n)}^{i_1, \ldots, i_{2n}} : \bigotimes_{k=1}^{2n} \vco{s}_{i_k}.
\end{equation}
Here, $\ten{J}_{(2n)}^{i_1, \ldots, i_{2n}}$ denotes a rank-$2n$ Cartesian interaction tensor coupling the spin sites $i_{1}, \ldots, i_{2n}$ (not necessarily distinct, but excluding duplicated summation). This representation establishes a systematic hierarchy in which magnetic interactions of increasing complexity are incorporated in a controlled and symmetry-consistent manner. The generalized Heisenberg Hamiltonian is recovered as the lowest-order member of this hierarchy, while biquadratic, three-spin, four-spin, and higher-order interactions arise naturally from successive tensor ranks.

Unlike conventional model Hamiltonians, which are typically constructed by selecting specific interaction terms based on physical intuition or prior knowledge, the tensorial formulation provides a complete and unbiased representation of the magnetic energy. All symmetry-allowed interactions are treated on an equal footing and can subsequently be identified, quantified, and interpreted directly from first-principles calculations without imposing a predefined model. This general formulation therefore provides the mathematical foundation for a first-principles methodology in which the relevant hierarchy of magnetic interactions is determined from the electronic structure rather than specified a priori. The automated construction and extraction of the resulting tensorial Hamiltonians are described in the following sections.

\subsection{Physical constraints defining the tensorial Hamiltonian}
\label{sec:physical_constraints}

\begin{figure*}[htbp] 
\centering
\includegraphics[width=\textwidth]{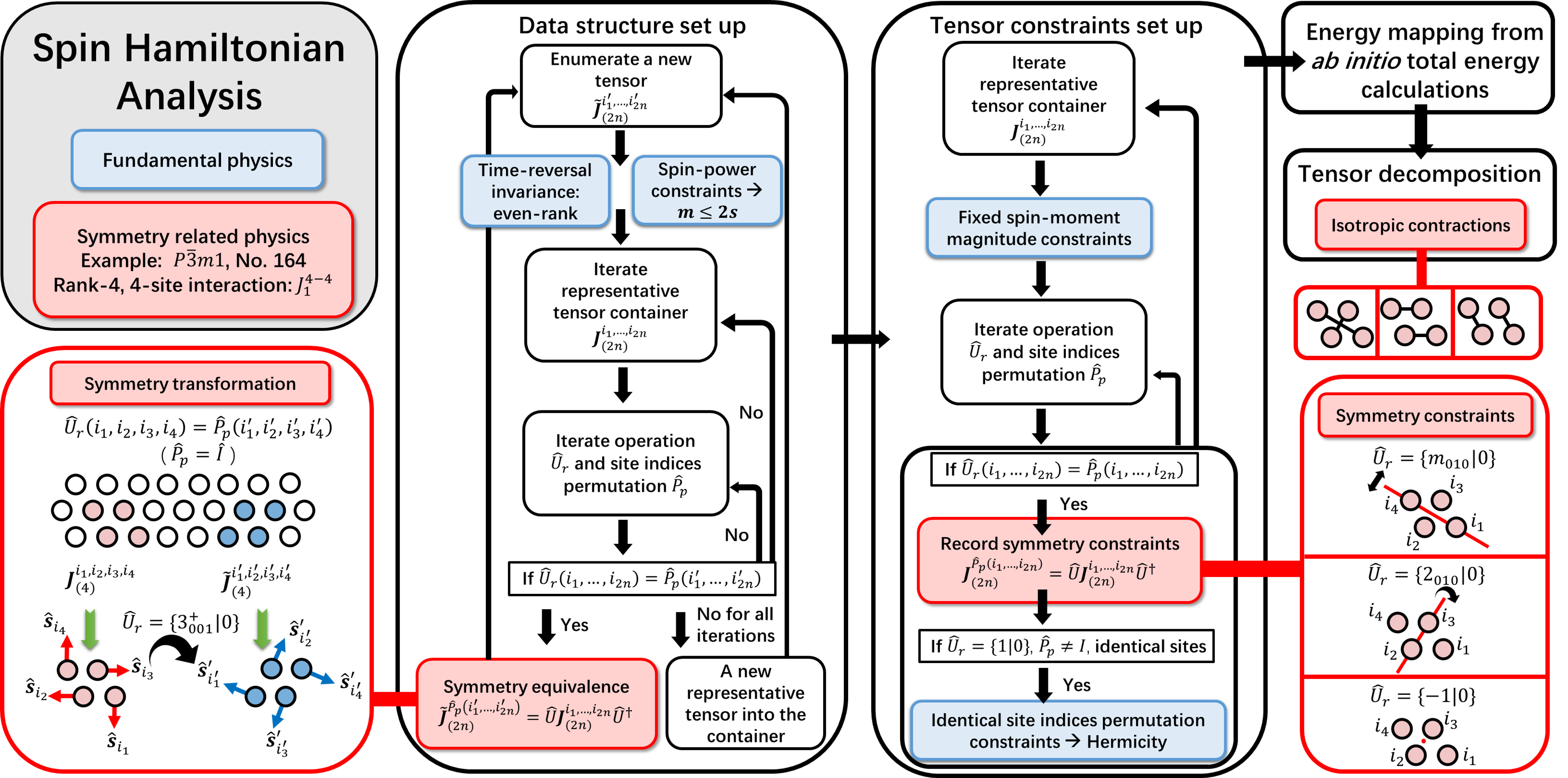} 
\caption{\textbf{Analysis of the spin Hamiltonian with tensorial interactions.} Schematic workflow illustrating the construction of the tensorial data structure and the implementation of constraints. The first rank-4, four-site interaction tensor is used as an example to demonstrate symmetry transformations, symmetry constraints, and isotropic contractions (marked in red) for a spin lattice with space group $P\bar{3}m1$ (No.\ 164). Constraints arising from fundamental physical principles are indicated in blue.}
\label{fig:spin-H-analysis}
\end{figure*}

While the tensorial expansion introduced above provides a formally complete representation of the magnetic energy, completeness alone does not guarantee a unique or physically meaningful Hamiltonian. Without additional constraints, the expansion would contain redundant interaction tensors that cannot be uniquely determined from first-principles calculations. The tensorial Hamiltonian is therefore subject to a small number of fundamental physical principles that eliminate these redundancies while preserving its complete descriptive power. Together, these constraints transform the formally complete expansion into a unique, non-redundant, and computationally tractable representation of the magnetic energy.

The first kind of fundamental constraint originates from time-reversal symmetry. In thermal equilibrium and in the absence of external magnetic fields, the Hamiltonian must remain invariant under time reversal. Because spin operators are odd under time reversal, only interaction terms containing an even number of spin operators are invariant. Consequently, all interaction tensors possess even rank, and the spin Hamiltonian is obtained as a scalar through complete contraction of all Cartesian spin indices.

The second kind of constraints arise from the local quantum spin algebra underlying Eq.~(\ref{eq:tensorial_spin_hamiltonian}): 1.\ Fixed spin-moment magnitude constraints; 2.\ Identical site indices permutation constraints; 3.\ Spin-power constraints. 

The magnitude of each local spin operator $\vco{s}_{i}$, or equivalently the magnetic moment, remains unchanged under magnetic excitations,
\begin{equation}
(\hat{s}_{i}^{x})^{2}+(\hat{s}_{i}^{y})^{2}+(\hat{s}_{i}^{z})^{2}=s_{i}(s_{i}+1)\,\hat{I},
\end{equation}
where $\hat{I}$ denotes the identity operator. This approximation is well justified for systems with localized magnetic moments, particularly insulating magnets with a finite band gap, where longitudinal spin fluctuations are energetically suppressed. Under this assumption, intra-site exchange contributions of the form $I_i(\vco{s}_i)^2$ are constant and are therefore omitted. The fixed spin-length constraint has an important consequence for the tensorial representation. Traces over Cartesian components associated with repeated site indices do not constitute independent physical interactions. Still, they can always be reduced to lower-rank interaction tensors or absorbed into the spin-independent reference energy. Tensorial interaction coefficients containing repeated site indices must therefore be traceless with respect to their Cartesian spin components, $ \mathrm{Tr}^{i}_{\alpha_i} \ten{J}_{(2n)}^{\ldots,i,\ldots,i,\ldots}=0$. A familiar example is the rank-2 single-ion anisotropy tensor (SIA), whose trace contributes only a constant energy shift and is therefore set to vanish.

Whenever identical lattice-site labels occur within an interaction tensor $\ten{J}_{(2n)}^{i_1,\ldots,i_{2n}}$, exchanging Cartesian indices associated with the same lattice site leaves the interaction invariant, giving rise to permutation symmetry relations between tensor components. The rank-2 single-ion anisotropy tensor, for example, is symmetric with respect to its Cartesian indices. Together with time-reversal symmetry, these permutation relations guarantee the Hermiticity of every interaction term in Eq.~(\ref{eq:tensorial_spin_hamiltonian}).

Finally, the finite dimensionality of the local Hilbert space imposes an upper bound on the spin order. For a spin quantum number $s_i$, operator products containing more than $2s_i$ powers of $\vco{s}_i$ are linearly dependent and can always be expressed in terms of lower-order operators. Consequently, the number of occurrences $m_i$ of spin operators associated with a given lattice site in any tensorial interaction must satisfy $m_i \leq 2s_i$, thereby ensuring a complete and non-redundant operator basis. For example, rank-2 single-ion anisotropy and biquadratic interactions are absent for $s=\tfrac{1}{2}$, while rank-4 single-ion anisotropy cannot occur for $s<2$.

\subsection{Constraints imposed by crystal symmetry}
\label{sec:crystal_symmetry}

While the preceding subsection established the universal physical constraints that define a unique and non-redundant tensorial Hamiltonian, the remaining interaction tensors are determined by the crystallographic symmetry of the underlying material. Space-group operations do not generate new magnetic interactions; instead, they identify tensor components that are symmetry equivalent and therefore describe the same physical interaction. Consequently, crystal symmetry further reduces the number of independent interaction parameters and transforms the general tensorial Hamiltonian into a compact, material-specific representation.

For magnets with localized magnetic moments, the effective spin Hamiltonian is assumed to be independent of the particular magnetic configuration. This approximation is well justified because the electronic states mediating the magnetic exchange preserve the symmetry of the underlying crystal potential. Consequently, the effective spin Hamiltonian inherits the complete space-group symmetry of the crystal, including spin--orbit coupling.

A general space-group operation is represented by the direct product $\hat{U}=\hat{U}_{r}\otimes\hat{U}_{s}$,  where $\hat{U}_{r}$ acts on the real-space coordinates through rotations, translations and inversions, while $\hat{U}_{s}$ denotes the corresponding SU(2) rotation acting on the spin degrees of freedom associated with the rotational part of $\hat{U}_{r}$. Spatial and spin rotations are therefore locked through spin--orbit coupling, ensuring that the tensorial Hamiltonian transforms covariantly under every symmetry operation.

A space-group operation generally maps one cluster of lattice sites, $(i_1,\ldots,i_{2n})$, onto a symmetry-equivalent cluster, $(i'_1,\ldots,i'_{2n})$,
\begin{equation}
\hat{U}_{r}(i_1,\ldots,i_{2n}) = \hat{P}_{p}(i'_1,\ldots,i'_{2n}),
\end{equation}
where $\hat{P}_{p}$ denotes the $p^{\mathrm{th}}$ permutation operator acting on the site labels. The associated interaction tensor transforms as a rank-$2n$ Cartesian tensor,
\begin{equation}
\tilde{\ten{J}}_{(2n)}^{\hat{P}_{p}(i'_1,\ldots,i'_{2n})} = \hat{U} \, \ten{J}_{(2n)}^{i_1,\ldots,i_{2n}}  \hat{U}^{\dagger}.
\end{equation}
This transformation establishes linear relations between tensor components belonging to symmetry-related clusters. Consequently, these tensor elements are not independent but are connected by the action of the space group. The symmetry operations therefore partition the complete tensor expansion into equivalence classes of interaction tensors, each represented by a single independent tensor. The resulting symmetry-adapted tensor representation preserves the complete physical content of the Hamiltonian while drastically reducing the number of independent interaction parameters.

An example is illustrated in Fig.~\ref{fig:spin-H-analysis} for the triangular two-dimensional lattice with space group $P\bar{3}m1$ and the rank-4 four-site interaction tensor $J_{1}^{4-4}$ (see the naming convention in Supplementary Note~1.2\cite{SupplementaryInfo}). The space-group operation $\hat{U}_{r}=\{3^{+}_{001}|0\}$, expressed in Seitz notation, corresponds to a clockwise $120^{\circ}$ rotation about the $z$ axis without translation. Since $\hat{P}_{p}=\hat{I}$, no relabelling of lattice sites occurs. The symmetry operation therefore rotates the red four-site cluster into the symmetry-equivalent blue cluster while simultaneously rotating the spin directions.

Additional constraints emerge whenever a space-group operation maps a cluster onto a permutation of itself,
\begin{equation}\label{eq:Ui=Pi}
\hat{U}_{r}(i_1,\ldots,i_{2n}) = \hat{P}_{p}(i_1,\ldots,i_{2n}),
\end{equation}
for which Hamiltonian invariance requires
\begin{equation} \label{eq:sym-cons}
\ten{J}_{(2n)}^{\hat{P}_{p}(i_1,\ldots,i_{2n})} = \hat{U} \,\ten{J}_{(2n)}^{i_1,\ldots,i_{2n}} \hat{U}^{\dagger}.
\end{equation}
Figure~\ref{fig:spin-H-analysis} illustrates three representative examples of such symmetry operations for the same lattice: mirror reflection in the $(010)$ plane, $\hat{U}_{r}=\{m_{010}|0\}$, two-fold rotation about the $010$ axis $\hat{U}_{r}=\{2_{010}|0\}$, and inversion, $\hat{U}_{r}=\{-1|0\}$. Each operation induces a corresponding permutation of the lattice sites, for example $\{-1|0\}(i_1,i_2,i_3,i_4)=(i_4,i_3,i_2,i_1)$.

A particularly important case is $\hat{U}_{r}=\hat{I}$, which is present in every space group. Equation~(\ref{eq:sym-cons}) then reduces to constraints arising solely from permutations. If $\hat{P}_{p}=\hat{I}$, the relation is trivially satisfied. Otherwise, repeated lattice-site labels must occur, recovering the identical site permutation constraints introduced in the previous subsection. A familiar example is the rank-2 single-ion anisotropy tensor, whose symmetry under exchange of Cartesian indices follows directly from this relation. Taken together, crystallographic symmetry transforms the complete tensor expansion into a symmetry-adapted tensor representation by grouping symmetry-equivalent interaction tensors into a single independent representative. This compact representation retains the full physical content of the Hamiltonian and forms the basis for the physically interpretable tensor decomposition introduced in the following subsection.

The complete set of universal physical constraints together with the crystallographic symmetry constraints is implemented algorithmically in \AMATIS, enabling the automatic construction of symmetry-adapted tensorial spin Hamiltonians. While the fundamental constraints are always enforced to guarantee a unique Hamiltonian, crystallographic symmetry can be selectively imposed, thereby allowing controlled investigations of symmetry-breaking phenomena.

To illustrate the practical impact of crystal symmetry, Table~\ref{tab:main1} summarizes the number of independent tensor components for three representative spin lattices. When only the universal physical constraints are imposed, interaction tensors without repeated lattice sites, such as rank-2 two-site and rank-4 four-site interactions, remain unconstrained. In contrast, the rank-2 single-ion anisotropy tensor already satisfies four independent constraints: one tracelessness condition originating from the fixed spin-length constraint and three permutation relations arising from identical lattice-site indices. Incorporating crystallographic symmetry further reduces the number of independent tensor components, with progressively stronger reductions observed for higher-symmetry lattices such as $P\bar{3}1m$ and $P\bar{3}m1$.

\begin{table}[h]
\centering
\caption{\textbf{Number of independent tensor components under different constraint levels:} (i) no constraint; (ii) fundamental constraints only; (iii) fundamental and symmetry constraints. For the space groups  $P\bar{3}1m$, $P\bar{3}m1$, $P3$, the numbers of independent rank-2 (rank-4) tensors are 6 (2), 9 (7), 10 (3), respectively.}
\label{tab:main1}
\begin{tabular}{lcccc}
\hline
Space group & Constraint type & Rank-2 & Rank-4 & Total \\ \hline
\multirow{3}{*}{$P\bar{3}1m$} & (i) & 54 & 162 & 216 \\ 
 & (ii) & 50 & 50 & 100 \\ 
 & (iii) & 26 & 24 & 50 \\ \hline
\multirow{3}{*}{$P\bar{3}m1$} & (i) & 81 & 567 & 648 \\ 
 & (ii) & 77 & 327 & 404 \\ 
 & (iii) & 37 & 106 & 143 \\ \hline
\multirow{3}{*}{$P3$} & (i) & 90 & 243 & 333 \\ 
 & (ii) & 82 & 75 & 157 \\ 
 & (iii) & 74 & 75 & 149 \\ \hline
\end{tabular}
\end{table}

\subsection{Tensor decomposition into physically interpretable interactions}
\label{sec:tensor_decomposition}

The symmetry-adapted tensor representation introduced above provides a complete and compact description of the magnetic Hamiltonian. However, the tensor coefficients themselves are not yet directly connected to the magnetic interactions commonly used to describe quantum magnets. To establish this connection, the interaction tensors are systematically decomposed into orthogonal tensor components corresponding to physically interpretable interaction channels. In this way, conventional magnetic interactions emerge naturally as projections of the complete tensor expansion rather than being introduced \textit{a priori}.

For rank-2 interaction tensors, Moriya~\cite{Moriya1960} demonstrated that every tensor can be uniquely decomposed into an isotropic scalar contribution (Heisenberg exchange), an antisymmetric component corresponding to the Dzyaloshinskii--Moriya interaction, and a traceless symmetric exchange tensor. The same principle extends naturally to interaction tensors of arbitrary even rank.

Among these contributions, the isotropic terms exist even in the absence of spin--orbit coupling and are typically the dominant interactions in magnetic materials. They therefore constitute the leading-order contribution that any effective spin Hamiltonian must accurately reproduce. The isotropic part of a rank-$2n$ interaction tensor is obtained by projecting the tensor onto the complete basis of $(2n-1)!!$ isotropic tensors $\ten{I}_{(2n)}$,
\begin{equation}\label{eq:iso-term-J}
\sum_{q=0}^{(2n-1)!!-1} \left[\frac{1}{3^{n}} \,\ten{J}^{i_{1}, \ldots, i_{2n}}_{(2n)}: \ten{I}_{(2n)}^{q}\right]\left[\ten{I}_{(2n)}^{q}:\bigotimes_{k=1}^{2n}\vco{s}_{i_k} \right].
\end{equation}
Here, $\ten{I}_{(2n)}^{q}$ denotes the $q^{\mathrm{th}}$ isotropic tensor constructed from tensor products of Kronecker delta symbols. The first contraction projects the interaction tensor onto the isotropic basis and yields the corresponding scalar coupling constants, whereas the second contraction generates the associated rotationally invariant spin operators.

For rank-2 tensors ($n=1$), the decomposition yields the familiar Heisenberg interaction through a single spin invariant, $\vco{s}_i\cdot\vco{s}_j$. For rank-4 tensors ($n=2$), three independent isotropic contractions exist, corresponding to the three possible products of spin scalar products. Depending on whether the interaction involves two, three, or four independent lattice sites, these invariants describe biquadratic exchange~\cite{Kittel1960,Thouless1965}, four-spin three-site interactions~\cite{Hoffmann2020}, or ring-exchange interactions~\cite{Takahashi1977}, respectively (see Supplementary Note~2.7~\cite{SupplementaryInfo}).

Equation~(\ref{eq:iso-term-J}) therefore provides the general projection of the complete tensor expansion onto its isotropic interaction channel for the cluster $(i_{1},\ldots,i_{2n})$ in Eq.~(\ref{eq:tensorial_spin_hamiltonian}). The identical-site permutation constraints discussed above guarantee that these projected interactions are Hermitian. Figure~\ref{fig:spin-H-analysis} illustrates the three isotropic contractions of the rank-4 interaction tensor $J_{1}^{4-4}$. Owing to the symmetry of the four-site cluster, the last two contractions are symmetry equivalent.

\subsection{Classical representation of the tensorial Hamiltonian}
\label{sec:classical_representation}

The complete tensor expansion, together with its symmetry-adapted tensor representation and physically interpretable decomposition, provides a general quantum-mechanical description of magnetic interactions. For most magnetic materials, however, thermodynamic properties, magnetic phase diagrams, and spin textures are obtained from atomistic spin simulations based on classical spin degrees of freedom. A classical representation of the tensorial Hamiltonian therefore provides the essential link between the formal tensor framework and large-scale materials simulations.

Because the fundamental physical constraints introduced above preserve the underlying quantum spin algebra, the tensorial quantum spin Hamiltonian can be mapped consistently onto a classical spin Hamiltonian in which three-dimensional unit vectors replace the spin operators. In practice, the interaction tensors are determined by energy mapping from first-principles calculations, where the total energies of a set of classical spin configurations are evaluated, and the corresponding tensorial interaction coefficients are extracted. Throughout this work, spins are therefore represented by unit vectors and the tensorial interaction parameters obtained with \AMATIS\ are expressed directly in units of energy (meV).

Whenever a quantum spin Hamiltonian is required, for example for quantum Monte Carlo simulations or other quantum many-body approaches, the classical interaction tensors can be transformed back into their quantum counterparts through the renormalization
\begin{equation}\label{eq:J_renorm}
\ten{J}_{(2n);\mathrm{Q}}^{i_{1},\ldots,i_{2n}} = \prod_{k=1}^{2n} \sqrt{\frac{s_{i_k}}{s_{i_k}+1}} \, \ten{J}_{(2n);\mathrm{Cl}}^{i_{1},\ldots,i_{2n}}, \end{equation}
where $s_{i_k}$ denotes the spin quantum number at lattice site $i_k$. This mapping preserves the correspondence between the classical and quantum descriptions while allowing the tensorial Hamiltonian to be employed consistently within both frameworks.

Further details on the derivation of the physical and crystallographic symmetry constraints, the tensor decomposition, and the implementation of the corresponding tensor contractions are provided in Supplementary Note~2~\cite{SupplementaryInfo}.

\subsection{Automated construction using \AMATIS}
\label{sec:automated_construction}

The theoretical framework developed above establishes three essential elements for an unbiased first-principles description of magnetic interactions: a complete tensor expansion, a symmetry-adapted tensor representation, and a physically interpretable tensor decomposition. A fourth important element is the choice of bias-free magnetic reference states. Applying this framework to realistic materials requires the systematic determination of the independent tensor coefficients from first-principles calculations. Owing to the large number of interaction clusters, tensor ranks, symmetry relations, and magnetic configurations involved, this construction rapidly becomes computationally intractable without dedicated automation.

We therefore implement the framework in \AMATIS\ (Automated Magnetism Analysis via Tensorial Interacting Spins), a computational framework designed to automatically construct tensorial spin Hamiltonians from first-principles calculations. Rather than assuming a predefined spin model, \AMATIS\ determines the symmetry-independent tensor interactions consistent with the physical and crystallographic constraints introduced above up to the chosen rank and interaction range and thus provides a converged tensorial Hamiltonian.

The resulting first-principles construction workflow is summarized in
Fig.~\ref{fig:workflow}. Starting from the crystal structure, a chosen maximum tensor rank and a predefined precision determining the interaction range, the framework first identifies all symmetry-independent interaction clusters and constructs the corresponding tensor basis by imposing the universal physical constraints and crystallographic symmetry relations described above. \AMATIS\ automates these operations and generates a minimal set of magnetic configurations to sample the complete interaction space. Their total energies, calculated using first-principles electronic-structure methods, are subsequently mapped onto the tensorial Hamiltonian, yielding the independent interaction tensors through an overdetermined constrained linear least-squares optimization. Finally, the reconstructed tensors are decomposed into physically interpretable interactions, providing direct access to conventional magnetic exchange parameters together with higher-order magnetic interactions that emerge naturally within the tensorial framework.

The resulting workflow is fully automated, scalable with tensor rank and interaction range, and independent of the underlying first-principles electronic-structure method. Consequently, the same methodology can be applied to a broad class of magnetic materials ranging from conventional Heisenberg magnets to systems exhibiting strong spin--orbit coupling, frustrated magnetism, or higher-order magnetic interactions.

\section*{Materials showcases}
\setcounter{subsection}{0}
To demonstrate the predictive capability of the framework, we apply it to representative 2D magnetic systems in which short- and long-range isotropic, relativistic anisotropic, and higher-order spin interactions can compete on comparable energy scales. Two-dimensional magnets and van der Waals heterostructures constitute particularly demanding test cases for a tensorial spin-Hamiltonian approach because their magnetic interactions can be tuned by interface engineering through substrates, strain and charge transfer, thereby substantially reshaping the magnetic energy landscape and stabilizing a rich spectrum of competing and emergent magnetic phases.

We consider two complementary classes of Cr-based systems, both characterized by spin moments of $3~\mu_{\mathrm{B}}$ (S=3/2) on the Cr sites. The first comprises  honeycomb CrI\textsubscript{3}-based insulators (Fig.~\ref{fig:CrI3}), suspended\cite{Xu2020,Huang2017,McGuire2015} and on 2H-WSe\textsubscript{2}\cite{Zhong2017} where substrate-induced symmetry breaking gives rise to additional  relativistic interactions generates chiral relativistic  interactions absent in the isolated monolayer.  The second comprises triangular CrTe\textsubscript{2}\cite{Meng2021,Zhang2021,Xian2022} and CrSe\textsubscript{2}\cite{Li2021,Wu2022} monolayers (Fig.~\ref{fig:CrXe2_iso_J}), which provide a prototypical platform for investigating frustrated itinerant magnetism driven by a whole spectrum of competing interactions that can be further tuned by lattice strain.\cite{Gong2019} Details of the computational procedure and full tensor components are provided in Supplementary Notes 5 and 8.\cite{SupplementaryInfo}

\subsection{Honeycomb CrI\textsubscript{3}: monolayer and heterostructure}

\begin{figure*}[htbp] 
\centering
\includegraphics[width=\textwidth]{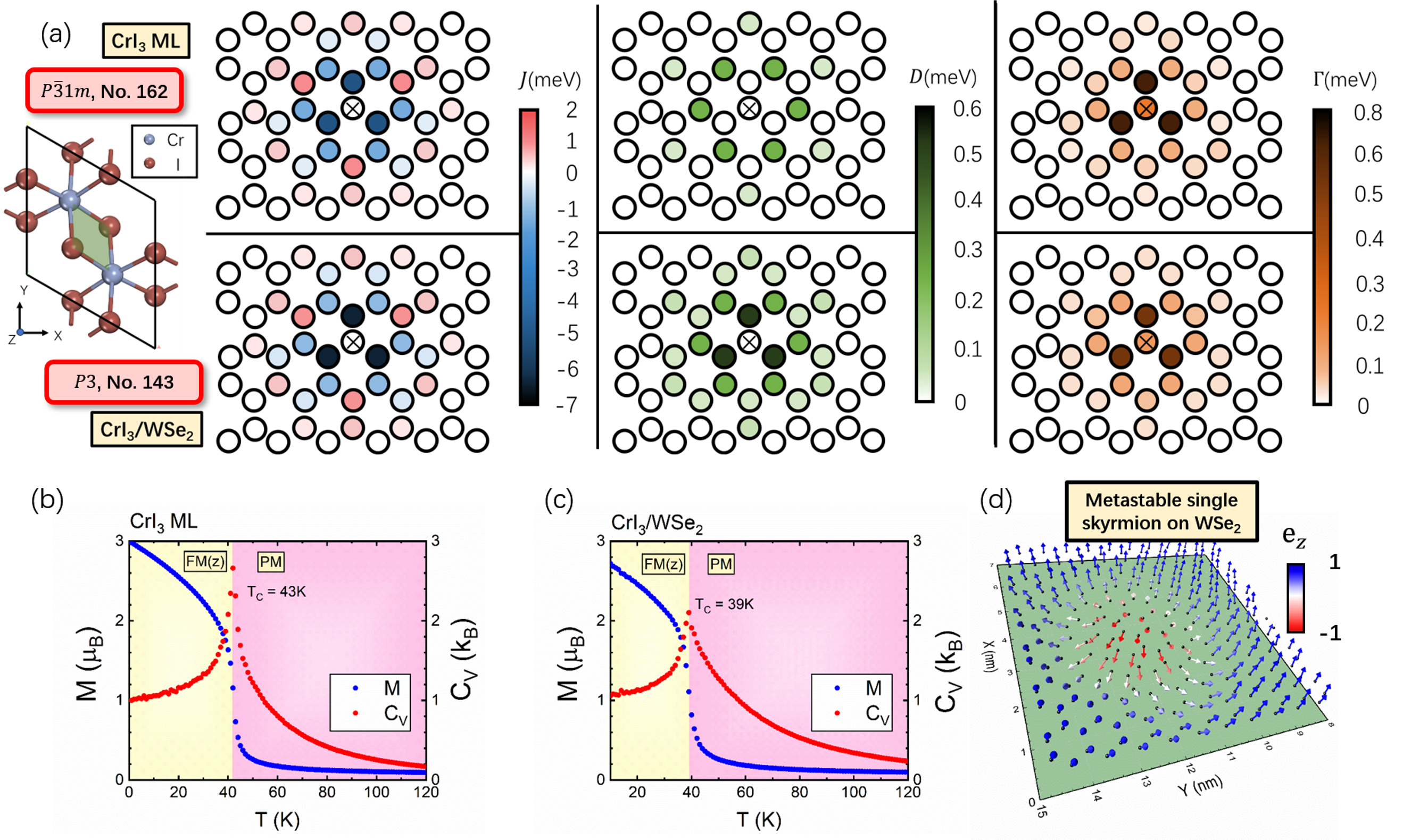} 
\caption{\textbf{Tensorial spin interactions and finite-temperature magnetism in CrI\textsubscript{3} systems.} Honeycomb CrI\textsubscript{3} monolayer structure with the unit cell outlined in black; blue and pink spheres denote Cr and I ions, respectively. 
(a) Spin lattices of the CrI\textsubscript{3} monolayer (top) and CrI\textsubscript{3}/2H-WSe\textsubscript{2} heterostructure (bottom), with color maps indicating the magnitude and sign of the tensorial spin interactions between the central site ($\otimes$) and neighboring atoms, including (left panels) the Heisenberg exchange $J$ ($J<0$ corresponds to ferromagnetic (FM) pair interactions, consistent with Eq.~\ref{eq:tensorial_spin_hamiltonian}), (middle panels)  magnitude $D$ of DMI vector, and (right panels) symmetric anisotropic exchange $\Gamma$ with SIA at $\otimes$.  The heterostructure breaks an in-plane symmetry and generates two inequivalent Cr Wyckoff positions, leading to a slight splitting of some tensor components \textit{e.g.}\ different SIA not shown here. For more details, see Supplementary Note~8.2. 
(b,c) Temperature dependence of the magnetization $M$ and heat capacity $C_\textrm{V}$ per Cr atom for the monolayer (b) and heterostructure (c), obtained from classical Monte Carlo simulations. The FM phase with out-of-plane magnetization (FM($z$), yellow background) and the paramagnetic phase (PM, magenta background) are indicated together with the Curie temperature $T_\textrm{C}$. (d) Metastable N\'eel-type skyrmion stabilized in the CrI\textsubscript{3}/WSe\textsubscript{2} heterostructure, with arrows colored by the out-of-plane spin component $\mathbf{e}_z$.}
\label{fig:CrI3}
\end{figure*}

\AMATIS~identifies dominant nearest-neighbor (n.n.), sizable second- (n.n.n.) and third-nearest-neighbor rank-2 tensorial interactions in the CrI\textsubscript{3} monolayer, including oscillatory ferromagnetic and antiferromagnetic isotropic Heisenberg exchange $J$ (Fig.~\ref{fig:CrI3}(a), left panel), and symmetric anisotropic exchange $\matr{\Gamma}$  (right panel). The extracted interactions confirm that monolayer CrI\textsubscript{3} is a prototypical two-dimensional tensorial anisotropic Ising ferromagnet, in which out-of-plane magnetic order is stabilized by spin–orbit-induced single-ion (SIA$\,=0.54$~meV/Cr atom) and additional n.n two-ion (TIA$\,=0.41$~meV/Cr atom) anisotropy. In addition to the two-ion anisotropy, the symmetric exchange $\matr{\Gamma}$ contains bond-dependent Kitaev-type interactions of $K=1.79$~meV. While these contributions vanish upon summation over the unit cell, they couple spin components within the Cr–I–Cr–I  bonding plane (green plaquette in the structure model) and locally modify the magnetic anisotropy landscape. 

Although the centrosymmetric lattice of the isolated monolayer suppresses a global Dzyaloshinskii-Moriya interaction (DMI), local inversion-symmetry breaking along bonds connecting n.n.n.\ Cr sites generates local antisymmetric exchange interactions of approximately $0.20$~meV. The corresponding DMI vectors lie  within the plane normal to the monolayer (middle panel) and favor a transverse spin twisting with a unique rotational sense  parallel to the components of the DMI vector in the monolayer plane.

Despite the weak van der Waals interfacial coupling, which induces only minor quantitative changes in the isotropic, single-ion anisotropy (SIA \(=0.30,\,0.39\)~meV), symmetric two-ion anisotropy (TIA \(=0.41\)~meV), and Kitaev interaction (\(K=1.69\)~meV), the situation changes qualitatively in the CrI\textsubscript{3}/WSe\textsubscript{2} heterostructure (Fig.~\ref{fig:CrI3}). There, the substrate breaks inversion symmetry and activates a sizeable global interfacial DMI (iDMI) between n.n.\ spin pairs of about $-0.5$~meV with a DMI-vector in the plane of the monolayer, visible for the n.n.\ spin pairs in Fig.~\ref{fig:CrI3}(a) (middle panel).

Classical Monte Carlo simulations based on the extracted tensorial Hamiltonian show little difference in the Curie temperature between the monolayer (43~K, Fig.~\ref{fig:CrI3}(b)) and the heterostructure (39~K, Fig.~\ref{fig:CrI3}(c)), consistent with the relatively weak modification of exchange interactions and anisotropies and in agreement with experiment\cite{Zhong2017}. The iDMI in the van der Waals heterostructure stabilizes at low temperatures metastable N\'eel-type skyrmions with counterclockwise handedness  (Fig.~\ref{fig:CrI3}(d)), with a characteristic diameter of 3~nm and the formation energy of 22.3~meV. These results illustrate how tensorial spin-Hamiltonian analysis with \AMATIS~can capture the impact of interface engineering in van der Waals heterostructures on relativistic magnetic couplings and emergent topological spin textures. 

\subsection{Triangular CrTe\textsubscript{2} and CrSe\textsubscript{2} monolayers}

\begin{figure*}[htbp] 
\centering
\includegraphics[width=\textwidth]{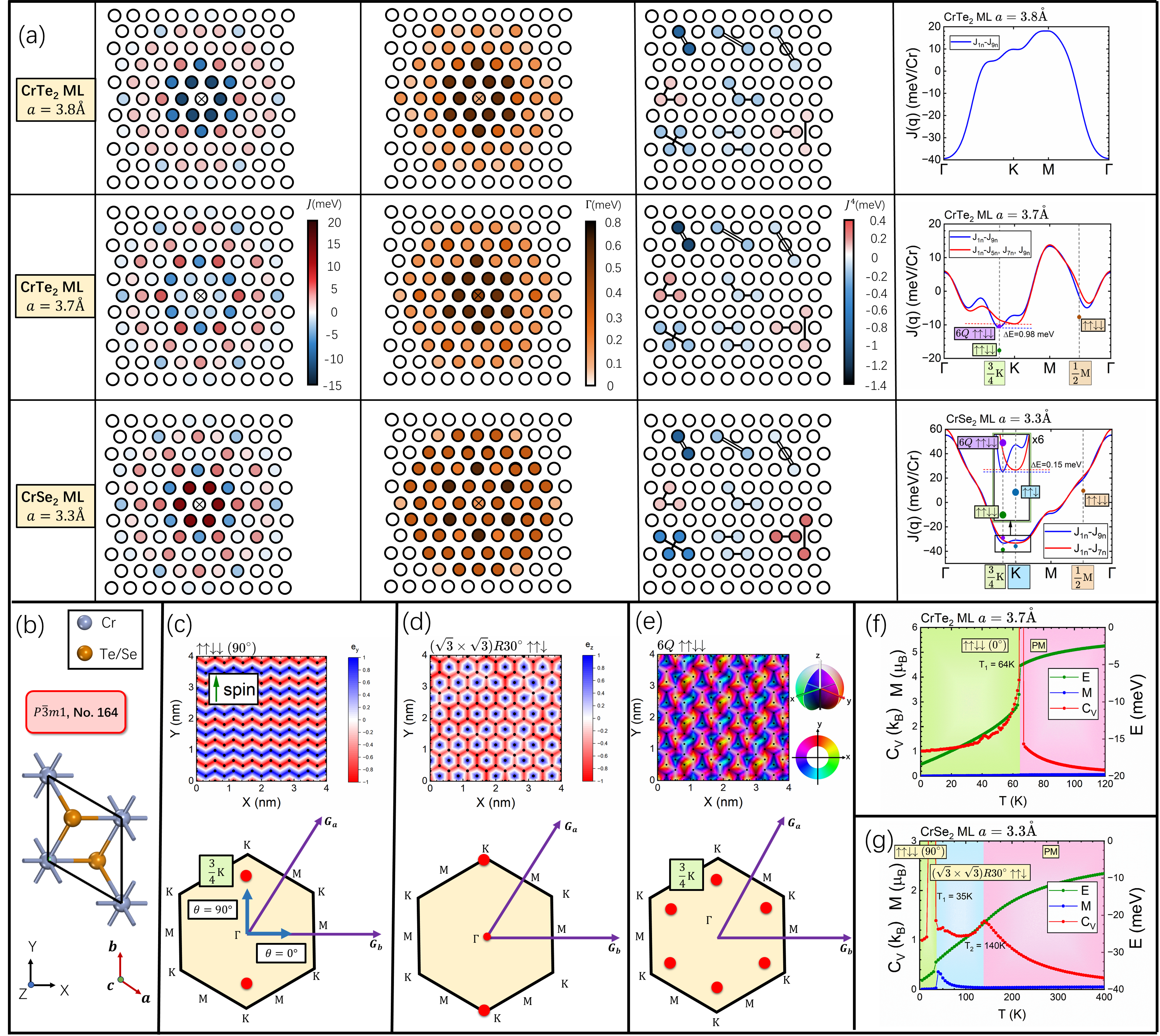} 
\caption{
\textbf{Evolution of tensorial spin interactions and magnetic phases in CrTe\textsubscript{2} and CrSe\textsubscript{2} monolayers.} (a) Real-space representation of the dominant tensorial spin interactions obtained from the first-principles tensorial construction for CrTe\textsubscript{2} with lattice constants $a=3.8$~\AA\ and $a=3.7$~\AA\, and for CrSe\textsubscript{2} with $a=3.3$~\AA\ (top to bottom). The reference Cr atom is marked by $\otimes$. The first three columns show the leading isotropic $J$ ($J<0$ denotes ferromagnetic coupling), two-ion (TIA) together with the single-ion anisotropy (SIA) at $\otimes$, and fourth-order spin interactions, while the right column displays the momentum-dependent isotropic exchange energy $J(\vc{q})$ (notice the different scales). CrTe\textsubscript{2} ($a=3.8$~\AA) exhibits a conventional ferromagnetic instability at the $\Gamma$-point. Upon reducing the lattice constant to $a=3.7$~\AA, the exchange interactions become strongly frustrated, producing low-energy spiral instabilities near $\mathbf q=3/4K$. A further reduction to CrSe\textsubscript{2} stabilizes antiferromagnetic order associated with the $K$-point of the hexagonal Brillouin zone. Energy differences between competing magnetic instabilities are indicated in the corresponding panels.
(b) Crystal structure of the monolayer transition-metal dichalcogenides and definition of the crystallographic and Cartesian coordinate systems.
(c-e) Three magnetic structures and their wave-vector $q$ representation in the first Brillouin zone.
(c) Collinear zigzag $(\uparrow\uparrow\downarrow\downarrow)$ ground state of CrSe\textsubscript{2} comprised of two wave vectors $\mathbf q_{\mathrm{zz}}=(0,\pm\pi)$ with in-plane easy axis perpendicular ($90^\circ$) to the zigzag chains.
(d) Ferrimagnetic $(\sqrt3\times\sqrt3)R30^\circ (\uparrow\uparrow\downarrow)$ state of CrSe\textsubscript{2}. The magnetic order corresponds to a three-sublattice $K$-point instability and carries a net out-of-plane magnetization.
(e) Non-coplanar $6Q$ state, a superposition of the three zigzag $(\uparrow\uparrow\downarrow\downarrow)$ states, with wave vectors $\mathbf q_{\mathrm{zz}}=3/4 K$ modulo 60$^\circ$ rotations.
Color code is to map the unity spin orientation vectors $\hat{\vc{e}}=(\phi,\theta)$ onto the unit sphere. 
(f,g) Finite-temperature Monte Carlo simulations based on the full tensorial spin Hamiltonian. Shown are the temperature dependence of the internal energy $E$, magnetization $M$, and heat capacity $C_V$, revealing the magnetic phase transitions ($T_1$) and ordering ($T_2$) temperatures. PM: paramagnetism.
}
\label{fig:CrXe2_iso_J}
\end{figure*}

We investigate monolayer CrTe\textsubscript{2} for the two experimentally reported lattice constants of 3.8~\AA\cite{Zhang2021,Meng2021} and 3.7~\AA,\cite{Xian2022} and compare them with monolayer CrSe\textsubscript{2} at its experimental lattice constant of 3.3~\AA.\cite{Liu2021} Figure~\ref{fig:CrXe2_iso_J} summarizes the extracted tensorial spin interactions (a), the resulting competing magnetic states and their ordering wave vectors (c-e), and the corresponding finite-temperature magnetic behavior obtained from Monte Carlo simulations (f,g).

For CrTe\textsubscript{2} at the larger lattice constant ($a=3.8$~\AA), \AMATIS\ predicts a conventional in-plane ferromagnetic ground state. The magnetism is dominated by a strong nearest-neighbor ferromagnetic Heisenberg exchange interaction ($-12.64$~meV), while a moderate in-plane single-ion anisotropy ($0.42$~meV/Cr) and symmetric anisotropic exchange interactions further stabilize the in-plane orientation. In this regime, the magnetic behavior is well described by a simple ferromagnetic spin model. 

A remarkably different picture emerges upon reducing the lattice constant by only $2.5$\% to $a=3.7$~\AA. The nearest-neighbor ferromagnetic exchange is suppressed by more than a factor of 20 in magnitude, and the isotropic exchange interactions acquire a pronounced long-range oscillatory character reminiscent of Ruderman--Kittel--Kasuya--Yosida (RKKY) interactions in metallic systems. As a consequence, the isotropic Heisenberg sector represented by the momentum-dependent isotropic exchange energy $J(\vc{q})$ no longer favors a ferromagnetic ground state but instead stabilizes a low-energy manifold of helical spin spirals characterized by a wave vector close to $\mathbf q=(0,\pi)=3/4K$ modulo the five additional symmetry-equivalent ones (Fig.~\ref{fig:CrXe2_iso_J}(c,e)), which is a period-four helix with a large $p$-wave altermagnetic spin-splitting of the electronic structure ($\max \Delta \epsilon_{\uparrow\downarrow}\approx 97$~meV).

Remarkably, \AMATIS\ reveals that the frustrated Heisenberg interactions alone do not determine the magnetic ground state. Instead, a sizeable nearest-neighbor biquadratic interaction ($-1.37$~meV) reconstructs the low-energy spiral manifold and stabilizes a commensurate collinear double-$Q$ ($2Q$) state ($\Delta E_{2Q-1Q}=-6.25$~meV/Cr) relative to the optimal single-$Q$ spiral. The resulting period-four up-up-down-down ($\uparrow\uparrow\downarrow\downarrow$) structure can be interpreted as the superposition of two counter-rotating helices with wave vectors $\mathbf q_{\mathrm{zz}}=(0,\pm\pi)$ ($\equiv \pm 1/4(1,1)$ in units of the reciprocal lattice) and a relative phase shift of $\pi/2$. In the terminology of two-dimensional magnetism, this corresponds to the zigzag phase (Fig.~\ref{fig:CrXe2_iso_J}(c)). The nearly identical stabilization energy obtained for the alternative row-wise $\uparrow\uparrow\downarrow\downarrow$ state at $\mathbf q=M/2$ ($\Delta E_{2Q-1Q}=-4.32$~meV/Cr, Fig.~\ref{fig:CrXe2_iso_J}(a)) suggests that  among the interactions originating from the rank-4 tensors,  the isotropic biquadratic and four-spin interaction provide the dominant mechanism for stabilizing period-four order, while the four-spin-three-site interaction contributes only marginally.

Concomitantly, the in-plane single-ion anisotropy increases by approximately 250\% to 1.08~meV/Cr atom. The symmetric anisotropic exchange tensor $\matr{\Gamma}$ further lifts the rotational degeneracy within the zigzag phase and selects an easy axis parallel to the zigzag chains, or equivalently perpendicular to the modulation vector $\mathbf q_{\mathrm{zz}}$ ($\uparrow\uparrow\downarrow\downarrow$~$(0^\circ)$ in Fig.~\ref{fig:CrXe2_iso_J}(c)).

The multi-site interactions control also the competition between the double- and sextuple-$Q$ ($6Q$) state (Fig.~\ref{fig:CrXe2_iso_J}(e), also Supplementary Fig.~12\cite{SupplementaryInfo}), a non-coplanar magnetic state with net-zero magnetic moment and a topological charge of $Q=-2$ per magnetic unit cell composed of a coherent superposition of all 3 symmetry equivalent double-$Q$ states oriented in three different orthogonal directions with an energy of $\Delta E_{6Q-2Q} = 7.29$~meV/Cr. The resulting finite real-space spin chirality renders this phase a potential source of an emergent magnetic field and a topological Hall response. 

For CrSe\textsubscript{2}, which possesses an even smaller lattice constant ($a=3.3$~\AA), the nearest-neighbor exchange becomes strongly antiferromagnetic ($17.56$~meV/Cr), favoring a $120^\circ$ N\'eel instability at the $K$-point of the triangular-lattice Brillouin zone. However, the long-range oscillatory exchange interactions place the system within approximately $0.15$~meV/Cr close to competing magnetic instabilities at $\mathbf q=(0,\pi)$. The combined action of frustrated exchange, an out-of-plane single-ion anisotropy ($-0.45$~meV/Cr), and anisotropic exchange interactions stabilizes a ferrimagnetic $(\sqrt3\times\sqrt3)R30^\circ$ up-up-down ($\uparrow\uparrow\downarrow$) state over the N\'eel state with a net out-of-plane magnetization of 3~$\mu_\text{B}$/magnetic unit cell (Fig.~\ref{fig:CrXe2_iso_J}(d)).  A sizeable nearest-neighbor biquadratic interaction ($-1.25$~meV/Cr) stabilizes the zigzag $2Q$- ($\Delta E_{2Q-1Q}=-7.70$~meV/Cr) and the alternative row-wise $2Q$-state ($\Delta E_{2Q-1Q}=-7.71$~meV/Cr) over the respective helical $1Q$-state,  and the $6Q$-state ($\Delta E_{6Q-2Q}=7.84$~meV/Cr) and changes the easy axis from out-of-plane to in-plane perpendicular to zigzag-chains ($\uparrow\uparrow\downarrow\downarrow$~$(90^\circ)$ in Fig.~\ref{fig:CrXe2_iso_J}(c)). The proximity of competing spiral, zigzag-like, and non-coplanar states demonstrates that CrSe\textsubscript{2}, like CrTe\textsubscript{2} at $a=3.7$~\AA, resides in a highly frustrated regime where long-range rank-2 and rank-4 tensorial interactions cooperate to determine the magnetic order.

The distinct magnetic ground states lead to markedly different finite-temperature behavior. Because the in-plane ferromagnetic state of CrTe\textsubscript{2} ($a=3.8$~\AA) retains the continuous rotational symmetry of an XY magnet, true long-range magnetic order is absent at finite temperature and a Berezinskii--Kosterlitz--Thouless transition is expected. In contrast, the zigzag ground states of CrTe\textsubscript{2} ($a=3.7$~\AA) and CrSe\textsubscript{2} break the sixfold in-plane degeneracy and therefore support long-range magnetic order. Monte Carlo simulations (Fig.~\ref{fig:CrXe2_iso_J}(f,g)) reveal a first-order transition from the zigzag phase to the paramagnetic state at $T_1=64$ K for CrTe\textsubscript{2} ($a=3.7$~\AA). For CrSe\textsubscript{2}, the zigzag phase first transforms into a ferrimagnetic $(\sqrt3\times\sqrt3)R30^\circ$ ($\uparrow\uparrow\downarrow$) phase at $T_1=35$ K, followed by a second-order transition into the paramagnetic state at $T_2=140$ K. The intermediate ferrimagnetic phase originates from the competition between in-plane and out-of-plane anisotropy contributions, as discussed in Supplementary Note~5\cite{SupplementaryInfo}.

\section*{Discussions}

The central outcome of this work is not the identification of a particular magnetic phase, but the demonstration that an unbiased tensorial representation of magnetic interactions can fundamentally change the construction and interpretation of spin Hamiltonians. In contrast to conventional approaches that begin from a predefined model and subsequently determine a limited set of parameters, the present framework derives the hierarchy of symmetry-allowed tensorial interactions directly from first-principles calculations. The resulting Hamiltonian therefore emerges from the electronic structure rather than from prior assumptions regarding the dominant magnetic mechanisms. Interactions of different rank, range, and symmetry are treated on equal footing, while the relevant hierarchy of tensorial couplings is determined by the accuracy required to reproduce the underlying first-principles magnetic energy landscape. AMATIS provides an automated realization of this methodology, enabling its application to realistic magnetic materials.

The material showcase highlights the consequences of this approach. While CrI\textsubscript{3} reproduces the established behavior of a prototypical van der Waals ferromagnet, CrI\textsubscript{3}/WSe\textsubscript{2} demonstrates how interface-induced symmetry breaking generates relativistic interactions that stabilize chiral magnetic textures absent in the pristine material. The chromium dichalcogenides reveal an even richer hierarchy of competing interactions, where isotropic exchange determines the ordering wave vector, higher-order interactions drive transitions between single-$Q$ and multi-$Q$ states, and anisotropic interactions select the spin orientation. The resulting magnetic landscape includes single-$Q$, double-$Q$, topologically nontrivial sextuple-$Q$, ferrimagnetic, and $p$-wave altermagnetic phases, as well as isolated skyrmionic textures, many of which would be difficult to anticipate within conventional spin-model approaches. The close energetic proximity of these states highlights the importance of capturing the full hierarchy of tensorial interactions, as small perturbations arising from growth conditions, strain, defects, or interfaces may readily tip the balance between competing magnetic orders.

More broadly, the framework establishes a general workflow for predictive Hamiltonian construction in magnetic materials. By combining symmetry analysis, tensor decomposition, constrained fitting, and automated post-processing within a unified methodology, it enables the systematic construction of transferable spin Hamiltonians suitable for thermodynamic, dynamical, and topological simulations. The ability to identify isotropic, anisotropic, and higher-order interactions on equal footing opens new opportunities for the discovery and design of emergent magnetic phases across diverse classes of quantum materials. \AMATIS{} provides an automated and scalable implementation of this workflow, offering a foundation for high-throughput exploration, inverse design, and data-driven discovery of magnetic materials and phenomena.

Finally, our results show that conventional magnetic interactions are  not assumed as the starting point of the construction, but emerge as physically interpretable projections of a symmetry-adapted tensorial representation derived from first-principles energies.

\section*{Methods}
\setcounter{subsection}{0}

\subsection*{Implementation of spin tensorial Hamiltonian construction}

\subsubsection{Data structures}

\begin{figure*}[htbp] 
\centering
\includegraphics[width=\textwidth]{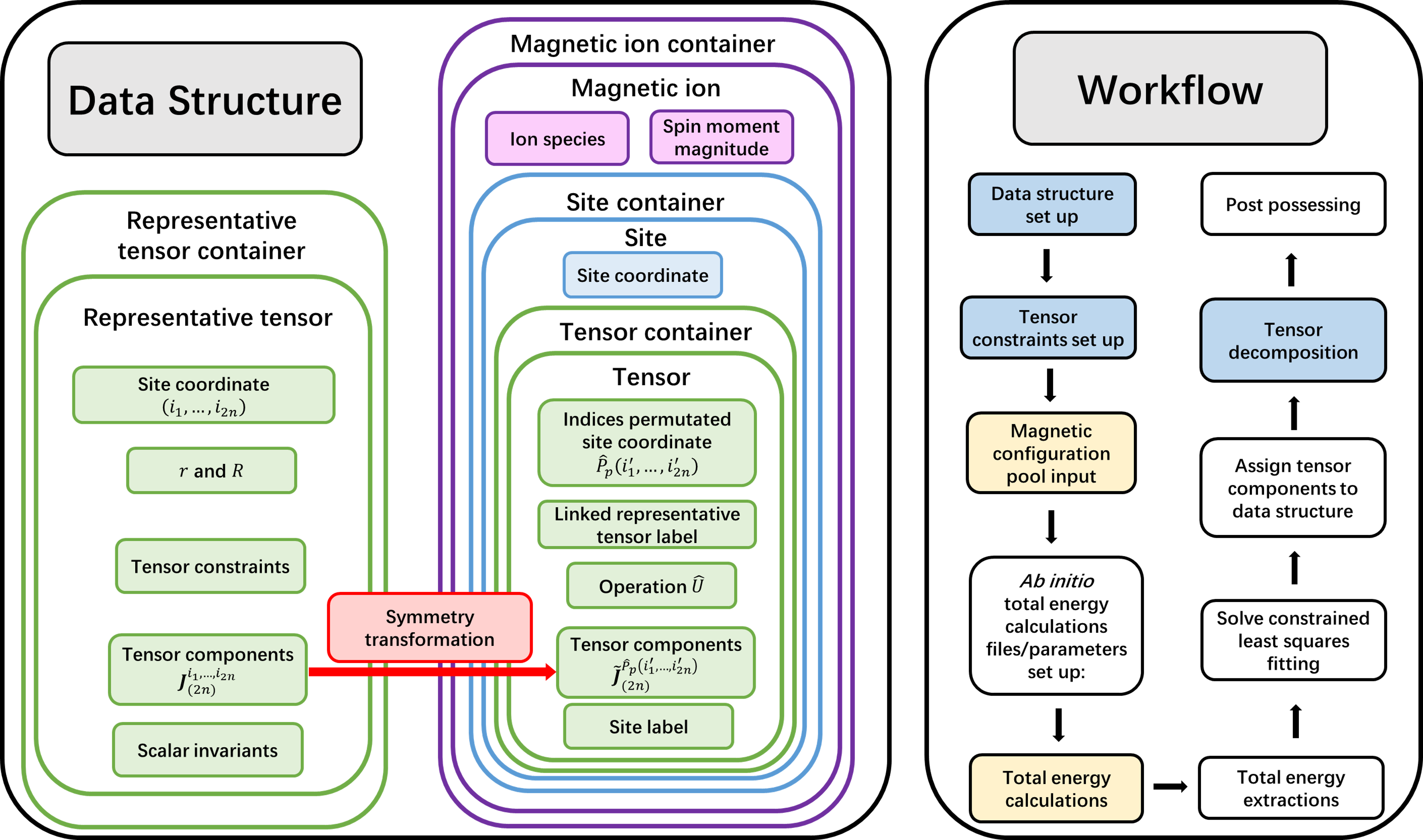} 
\caption{\textbf{First-principles construction of tensorial spin Hamiltonians.}
Left panel: Tensorial data structure implemented in \AMATIS{} for representing spin lattices, magnetic ions, and tensorial interactions. Hierarchical containers group related entities, with elements at the same level indicated by a common color. Tensor components of representative and site-resolved tensors are related through symmetry transformations (see Fig.~\ref{fig:spin-H-analysis}). Right panel: Automated workflow for constructing tensorial spin Hamiltonians from first-principles total energies. Blue blocks correspond to modules described in Fig.~\ref{fig:spin-H-analysis}, yellow blocks denote external modules interfaced with \AMATIS, and white and blue blocks represent fully automated procedures within \AMATIS.}
\label{fig:workflow}
\end{figure*}

We describe the implementation of the tensorial Hamiltonian
construction illustrated in Fig.~\ref{fig:workflow}. The underlying data structure represents magnetic ions and their tensorial interactions and provides the basis for applying symmetry and physical constraints. The implementation is based on an object-oriented hierarchy. Within the data structure, the container groups related objects. 

The ``Magnetic ion container'' groups the  ``Magnetic ion'' objects. "Magnetic ion" is a data block that stores the information of a magnetic ion element, the associated spin magnitude, and the "Site container" that groups symmetry-equivalent sites sharing the same Wyckoff position. "Site" is a data block that stores the spatial coordinate and the "Tensor container" that groups all tensors related to that site. "Tensor" is a data block that stores the necessary information for the site-related tensor, such as tensor components.

Tensor objects represent the interaction parameters associated with a given site. To avoid redundancy, the data block "Representative tensor"  is introduced for each symmetry-equivalent class of interactions, forming a minimal set of independent parameters for subsequent constraint enforcement and fitting. Representative tensors are linked to site-resolved tensors through symmetry transformations.

\subsubsection{Workflow}

Here, we outline the stepwise workflow of \AMATIS, corresponding to the right panel of Fig.~\ref{fig:workflow}, from structure input to post-processing. Implementation details are provided in Supplementary Note~3:\cite{SupplementaryInfo}

\begin{enumerate}
\def\labelenumi{\arabic{enumi}.}
\item
  Read the geometrically relaxed input structure, identify the magnetic ions, and the magnitude of the magnetic moments or spins, respectively.
\item
  Classify magnetic ions by their Wyckoff positions and construct the magnetic ion container shown in Fig.~\ref{fig:workflow}(left). 
\item
  Set up the tensors subject to constraints by initiating the workflows for the construction of the site and representative tensor containers. Establish symmetry relations to site-resolved tensors.  This process corresponds to the "Data structure set up" part in both Fig.~\ref{fig:spin-H-analysis} and~\ref{fig:workflow}.
\item
  Apply symmetry constraints to the representative tensors if symmetry constraints are switched on. Even if it is off, the identical site indices permutation constraints should be applied. Record all constraints in the "Tensor constraint" block in Fig.~\ref{fig:workflow}. This process corresponds to the "Tensor constraint set up" part in both Fig.~\ref{fig:spin-H-analysis} and~\ref{fig:workflow}.
\item
  Read selected magnetic configurations forming the configuration pool used to fit the tensor components (see Section~\ref{ssec:MCS} for details).
\item  
  Generate input files for first-principles (DFT) total-energy calculations.
\item
  Perform total-energy calculations in a high-throughput and parallelized manner.
\item
  Extract total energies from the calculations.
\item
  Determine tensor components of the representative tensors via constrained least-squares fitting.\cite{Boyd2018}  The resulting parameters are mapped to site-resolved tensors through symmetry transformations shown in Fig.~\ref{fig:workflow}.
\item
  Decompose the tensors and extract the isotropic invariants. This process corresponds to the "Tensor decomposition" part in both Fig.~\ref{fig:spin-H-analysis} and~\ref{fig:workflow}.
\item
  Transfer the resulting spin Hamiltonian to the post-processing modules, such as Monte Carlo simulations,  micromagnetic modeling, atomistic spin dynamics, or spin-wave analysis. In this work, thermodynamic properties are evaluated using classical and semiclassical Monte Carlo simulations (see Supplementary Note~7\cite{SupplementaryInfo}).
\end{enumerate}

\subsubsection{Magnetic configuration selection}\label{ssec:MCS}
To determine the tensorial interaction parameters from first-principles calculations, total energies are evaluated for a set of distinct spin configurations. The choice of configurations is essential to ensure a well-conditioned, rapidly convergent, and physically meaningful mapping onto the spin Hamiltonian while avoiding reference state bias. In particular, the selected configurations must sufficiently sample the relevant configuration space to resolve independent tensor components and minimize linear dependencies in the fitting procedure, enabling accurate extraction of both isotropic and anisotropic interactions. In the present framework, the relevant configuration space corresponds to the global manifold of fixed spin moments. A wide range of magnetic configurations can satisfy these requirements, including spin spirals in momentum space, magnetic states compatible with magnetic space groups commuting with the electronic Hamiltonian, and configurations approximating paramagnetic disorder.

In this work, the configuration pool consists of spin-spiral and symmetry-distinct magnetic states spanning different magnetic space groups. This strategy ensures broad coverage of spin orientations and correlation patterns.  The number of configurations required depends on several factors, including the spatial interaction range, lattice symmetry, the number of independent tensor coefficients, and the metallic or insulating character of the system. For the materials considered here, approximately 100-200 distinct configurations were sufficient to obtain converged tensor parameters with residual fitting errors in the range of $\sigma_E=0.10-0.35$~meV/unit cell. They can be either selected by hand or automatically enumerated via the preprocessing software\cite{Wang2026SpinGraph}. Additional details on the construction and validation of the configuration set are provided in Supplementary Note~6.\cite{SupplementaryInfo}

\subsubsection{Mapping to tensorial spin Hamiltonians}

The total energies obtained from DFT with spin-orbit coupling were mapped onto a general tensorial spin Hamiltonian using the \AMATIS~framework. The Hamiltonian is expressed as a sum of rank-2 and higher-rank spin-interaction tensors constrained by time-reversal symmetry and the crystallographic symmetries of the underlying material, including the spin--orbit-induced coupling between spatial and spin rotations. 

The mapping procedure determines the tensor components by minimizing the least-squares error between DFT energies and model energies over the full configuration set. Only symmetry-allowed tensor components were included. Rank-2 (bilinear) and rank-4 (biquadratic and multi-spin) interactions were considered, while higher-rank tensors were neglected due to their negligible contribution within the numerical accuracy.

The quality of the fitting was quantified by the unbiased standard error $\sigma_E$. Systematic convergence tests with respect to the number of magnetic configurations and the interaction range confirmed the robustness of the extracted tensorial spin Hamiltonians. The energy differences $\Delta E$ between competing single-$Q$ and multi-$Q$ states obtained from the fitted Hamiltonians agree with the corresponding first-principles values, provided in the main text, within $\sigma_E$, demonstrating the predictive accuracy of the approach. Detailed tensor components and convergence analyses are provided in the Supplementary Notes 8 and 5, respectively.\cite{SupplementaryInfo}

\subsection*{Density functional theory calculations}
\noindent\textsl{Core DFT Setup:} The \emph{ab initio} calculations are performed using the Vienna ab initio simulation package (VASP),\cite{Kresse1996} employing the projector augmented wave (PAW) method.\cite{Blochl1994} The orbitals $3s,\ 3p$ of Cl, $3p,\ 3d,\ 4s$ of Cr, $4s,\ 4p$ of Se, $4s,\ 4p$ of Br, $5s,\ 5p$ of Te, $5s,\ 5p$ of I, and $5p,\ 5d,\ 6s$ of W are taken into account in the pseudopotentials. The cutoff energies are 520~eV for CrX\textsubscript{3} honeycomb monolayers (ML) and 400~eV for 1T CrTe\textsubscript{2} and CrSe\textsubscript{2} ML.  K-point meshes of $2\times4\times1$ and $4\times8\times1$ were used for insulating and metallic systems, respectively, consistent with the corresponding supercell sizes. The electronic energy convergence criterion was set to $10^{-7}$~eV, and spin–orbit coupling was included. Experimental lattice constants were used for the two-dimensional magnetic systems.

\noindent\textsl{Exchange–Correlation and Hubbard $U$:}   The exchange-correlation functional is treated using the local spin density approximation (LSDA)\cite{Vosko:1980}  with a Hubbard $U=0.5$~eV on the Cr $d$ orbitals for CrCl\textsubscript{3}, CrBr\textsubscript{3}, CrI\textsubscript{3}, and CrI\textsubscript{3}/WSe\textsubscript{2}, following \onlinecite{Xu2018,Xu2020a}. For CrTe\textsubscript{2}, the Perdew--Burke--Ernzerhof (PBE) functional\cite{Perdew1996} with $U=1.5$~eV was used. For CrSe\textsubscript{2}, PBE with $U=2.5$~eV was used. These choices reproduce experimental transition temperatures and yield $S=3/2$ for Cr.

\textsl{Structural Setup:} The length of the lattice vector $c$  was set to 30~\AA, providing a vacuum spacing exceeding 20~\AA\ to avoid spurious interlayer interactions. Lattice constants were taken from experiment: CrCl\textsubscript{3} (6.00~\AA\cite{Lu2023}), CrBr\textsubscript{3} (6.38~\AA\cite{Huang2024}), CrI\textsubscript{3} (6.72~\AA\cite{Huang2017,Xu2020a}), CrTe\textsubscript{2} (3.70~\AA\cite{Xian2022} and 3.80~\AA\cite{Zhang2021,Meng2021,Sun2021}), and CrSe\textsubscript{2} (3.30~\AA\cite{Liu2021}). Atomic positions were relaxed until residual forces were below 0.03~eV/\AA. 

\textsl{Heterostructure Treatment:} For the CrI\textsubscript{3}/2H-WSe\textsubscript{2} heterostructure, van der Waals interactions were included using the DFT-D3 correction scheme.\cite{Grimme2010} To investigate the influence of the interface and substrate on the magnetic interactions while avoiding any strain in CrI\textsubscript{3}, a commensurate heterostructure was constructed using the lattice constant of monolayer CrI\textsubscript{3}, such that one CrI\textsubscript{3} unit cell matches a \(2\times2\) WSe\textsubscript{2} supercell, with the WSe\textsubscript{2} layer correspondingly adapted by less than 1\%.

\textsl{Supercells \& Magnetic Configurations:} Final supercells of $4\times2\times1$ were used for insulating systems, and $8\times4\times1$ for metallic systems. A large set of non-collinear magnetic configurations was generated using constrained magnetic moments~\cite{Ma2015} to enable accurate energy mapping.

The computational parameters are summarized in Supplementary Table~1. See Supplementary Note~4 for convergence tests and detailed parameters.\cite{SupplementaryInfo}

\subsection*{Spin-orbit-induced relativistic interactions}

Relativistic exchange interactions, including Dzyaloshinskii-Moriya and higher-order chiral and nematic terms, were extracted from the antisymmetric and anisotropic components of the tensor Hamiltonians. Their symmetry properties were analyzed using magnetic space-group constraints.

The magnitude and spatial range of relativistic interactions were evaluated for different materials, substrates, and strain conditions. This allowed a systematic comparison of spin-orbit-driven effects across distinct crystal environments.

\subsection*{Visualization of tensorial spin interactions}

To facilitate the interpretation of the extracted tensorial spin Hamiltonians, the dominant interactions are projected onto physically intuitive quantities and visualized in real and momentum space. Color maps show the magnitude and sign of spin interactions between the reference Cr atom marked by $\otimes$ and surrounding sites. The isotropic Heisenberg exchange interactions $J$ are obtained from the rank-2 tensors according to Eq.~(\ref{eq:iso-term-J}). Antisymmetric and symmetric anisotropic exchange interactions are represented by the traceless tensor $\ten{D}$ and $\ten{\Gamma}$, while the on-site contribution corresponds to the single-ion anisotropy. Their norms are extracted as $D$ and $\Gamma$. Fourth-order exchange interactions $J^4$ are obtained from the rank-4 tensors and projected onto effective biquadratic, three-site four-spin, and four-site ring-exchange channels according to their tensor contraction topology, which is visualized by the bond multiplicity among atom clusters. The momentum-dependent exchange energy $J(\mathbf q)=\frac{1}{2}\sum_{j}J_{j0}\expo{-i\vc{q}\cdot\vc{r}_{j0}}$ is calculated as the lattice Fourier transform of the isotropic exchange interactions between the center atom at $\vc{r}_0$ and the respective neighboring atoms at $\vc{r}_j$.  It is used to identify the dominant magnetic instabilities and their associated ordering wave vectors.

\subsection*{Color code for spin-orientations of the 3D sextuple-$Q$ state}
To visualize the non-coplanar $6Q$ state, we map the unit spin orientation vector $\hat{\vc{e}}$ to the unit sphere surface: the azimuthal angle of $\hat{\vc{e}}$ corresponds to the hue; the polar angle determines the color saturation. The colors at the equator are: green ($+x$); red ($+y$); violet ($-x$); cyan ($-y$).

\subsection*{Monte Carlo simulations}

Thermodynamic properties were computed using classical and semiclassical Monte Carlo (MC) simulations based on the  extracted spin Hamiltonians augmented by dipole--dipole interaction. 
Parallel tempering\cite{Hukushima1996} was employed to improve sampling efficiency in complex magnetic energy landscapes.

For CrI\textsubscript{3} and the CrI\textsubscript{3}/WSe\textsubscript{2} heterostructure, a $30 \times 30 \times 1$ lattice was used, while for CrTe\textsubscript{2} and  CrSe\textsubscript{2} a $36 \times 36 \times 1$ lattice was adopted to accommodate the  relevant magnetic superstructures. 

Each simulation was equilibrated for at least $5 \times 10^{4}$ Monte Carlo steps.  After thermal annealing, energy and magnetization, along with their second moments, were sampled for statistical averaging. Configuration exchanges between adjacent temperatures were attempted  every 10 steps to avoid trapping in local energy minima and to accelerate convergence.

The total number of Monte Carlo steps required to reach thermal equilibrium depends on the complexity of the energy landscape and was chosen such that all thermodynamic observables were well converged.

Magnetization, susceptibility, energy, and heat capacity were computed from ensemble averages. Phase transitions were identified from peaks in the susceptibility and heat capacity. Additional details on the simulation protocol and convergence tests are provided in the  Supplementary Note~7.\cite{SupplementaryInfo}

\textbf{\hfill\break}

\section*{References}
\bibliographystyle{naturemag}
\bibliography{amatis.bib}

@article{Heisenberg1928,
  author  = {Heisenberg, W.},
  title   = {On the theory of ferromagnetism},
  journal = {Z. Phys.},
  volume  = {49},
  pages   = {619--636},
  year    = {1928},
  doi     = {10.1007/BF01328601}
}

@article{Anderson1950,
  author  = {Anderson, P. W.},
  title   = {Antiferromagnetism. \uppercase{T}heory of superexchange interaction},
  journal = {Phys. Rev.},
  volume  = {79},
  pages   = {350--356},
  year    = {1950},
  doi     = {10.1103/PhysRev.79.350}
}

@article{Kittel1960,
  author  = {Kittel, C.},
  title   = {Model of exchange-inversion magnetization},
  journal = {Phys. Rev.},
  volume  = {120},
  pages   = {335--342},
  year    = {1960},
  doi     = {10.1103/PhysRev.120.335}
}

@article{Thouless1965,
  author  = {Thouless, D. J.},
  title   = {Exchange in solid $^3$\uppercase{H}e and the \uppercase{H}eisenberg \uppercase{H}amiltonian},
  journal = {Proc. Phys. Soc.},
  volume  = {86},
  pages   = {893--904},
  year    = {1965},
  doi     = {10.1088/0370-1328/86/5/310}
}

@article{Takahashi1977,
  author  = {Takahashi, M.},
  title   = {Half-filled \uppercase{H}ubbard model at low temperature},
  journal = {J. Phys. C: Solid State Phys.},
  volume  = {10},
  pages   = {1289--1300},
  year    = {1977},
  doi     = {10.1088/0022-3719/10/8/014}
}

@article{Hoffmann2020,
  author  = {Hoffmann, M. and Blügel, S.},
  title   = {Systematic derivation of realistic spin models for beyond-\uppercase{H}eisenberg solids},
  journal = {Phys. Rev. B},
  volume  = {101},
  pages   = {024418},
  year    = {2020},
  doi     = {10.1103/PhysRevB.101.024418}
}

@article{Grytsiuk2020,
  author  = {Grytsiuk, S. and Hanke, J. P. and Hoffmann, M. and Bouaziz, J. and Gomonay, O. and Bihlmayer, G. and Lounis, S. and Mokrousov, Y. and Blügel, S.},
  title   = {Topological--chiral magnetic interactions driven by emergent orbital magnetism},
  journal = {Nat. Commun.},
  volume  = {11},
  pages   = {511},
  year    = {2020},
  doi     = {10.1038/s41467-020-14370-0}
}

@article{Gong2019,
  author  = {Gong, C. and Zhang, X.},
  title   = {Two-dimensional magnetic crystals and emergent heterostructure devices},
  journal = {Science},
  volume  = {363},
  pages   = {706},
  year    = {2019},
  doi     = {10.1126/science.aav4450}
}

@article{Zhong2017,
  author  = {Zhong, D. and Seyler, K. L. and Linpeng, X. and Cheng, R. and Sivadas, N. and Huang, B. andSchmidgall, E. and Taniguchi, T. and Watanabe, K. and McGuire, M. A. and Yao, W. and Xiao, D. and Fu, K. C. and Xu, X.},
  title   = {Van der \uppercase{W}aals engineering of ferromagnetic semiconductor heterostructures for spin and valleytronics},
  journal = {Sci. Adv.},
  volume  = {3},
  pages   = {e1603113},
  year    = {2017},
  doi     = {10.1126/sciadv.1603113}
}

@article{Drautz2005,
  author  = {Drautz, R. and Fähnle, M.},
  title   = {Parametrization of the magnetic energy at the atomic level},
  journal = {Phys. Rev. B},
  volume  = {72},
  pages   = {212405},
  year    = {2005},
  doi     = {10.1103/PhysRevB.72.212405}
}

@article{Drautz2004,
  author  = {Drautz, R. and Fähnle, M.},
  title   = {Spin-cluster expansion: \uppercase{P}arametrization of the general adiabatic magnetic energy surface with ab initio accuracy},
  journal = {Phys. Rev. B},
  volume  = {69},
  pages   = {104404},
  year    = {2004},
  doi     = {DOI: 10.1103/PhysRevB.69.104404}
}

@article{Bouaziz2025,
  title = {Spin models and cluster multipole method: Application to kagome magnets},
  author = {Bouaziz, Juba and Nomoto, Takuya and Arita, Ryotaro},
  journal = {Phys. Rev. B},
  volume = {112},
  issue = {1},
  pages = {014406},
  numpages = {17},
  year = {2025},
  month = {Jul},
  publisher = {American Physical Society},
  doi = {10.1103/m4wc-hhc4},
  url = {https://link.aps.org/doi/10.1103/m4wc-hhc4}
}

@article{Pindor1983,
doi = {10.1088/0305-4608/13/5/012},
url = {https://doi.org/10.1088/0305-4608/13/5/012},
year = {1983},
month = {may},
publisher = {},
volume = {13},
number = {5},
pages = {979},
author = {A J Pindor and J Staunton and G M Stocks and H Winter},
title = {Disordered local moment state of magnetic transition metals: a self-consistent \uppercase{KKR} \uppercase{CPA} calculation},
journal = {Journal of Physics F: Metal Physics}
}

@article{Szunyogh:2011,
  title = {Atomistic spin model based on a spin-cluster expansion technique: \uppercase{A}pplication to the \uppercase{I}r\uppercase{M}n${}_{3}$/\uppercase{C}o interface},
  author = {Szunyogh, L. and Udvardi, L. and Jackson, J. and Nowak, U. and Chantrell, R.},
  journal = {Phys. Rev. B},
  volume = {83},
  issue = {2},
  pages = {024401},
  numpages = {9},
  year = {2011},
  month = {Jan},
  publisher = {American Physical Society},
  doi = {10.1103/PhysRevB.83.024401},
  url = {https://link.aps.org/doi/10.1103/PhysRevB.83.024401}
}

@article{Hatanaka:2025,
author = {Hatanaka ,Tatsuto and Bouaziz ,Juba and Nomoto ,Takuya and Arita ,Ryotaro},
title = {Calculation of the \uppercase{B}iquadratic \uppercase{S}pin \uppercase{I}nteractions \uppercase{B}ased on the \uppercase{S}pin \uppercase{C}luster \uppercase{E}xpansion for \uppercase{A}b initio \uppercase{T}ight-binding \uppercase{M}odels},
journal = {Journal of the Physical Society of Japan},
volume = {94},
number = {12},
pages = {124709},
year = {2025},
doi = {10.7566/JPSJ.94.124709},
URL = { https://doi.org/10.7566/JPSJ.94.124709},
eprint = {https://doi.org/10.7566/JPSJ.94.124709}
}

@article{Moriya1960,
  author  = {Moriya, T.},
  title   = {Anisotropic superexchange interaction and weak ferromagnetism},
  journal = {Phys. Rev.},
  volume  = {120},
  pages   = {91--98},
  year    = {1960},
  doi     = {10.1103/PhysRev.120.91}
}

@article{Dzyaloshinskii1957,
  author  = {Dzyaloshinskii, I. E.},
  title   = {Thermodynamic theory of “weak” ferromagnetism in antiferromagnetic substances},
  journal = {Sov. Phys. JETP},
  volume  = {5},
  pages   = {1259--1272},
  year    = {1957}
}

@article{Brinker2019,
  author  = {Brinker, S. and dos Santos Dias, M. and Lounis, S.},
  title   = {The chiral biquadratic pair interaction},
  journal = {New J. Phys.},
  volume  = {21},
  pages   = {083015},
  year    = {2019},
  doi     = {10.1088/1367-2630/ab2f3f}
}

@article{Laszloffy2019,
  author  = {Lászlóffy, A. and Rózsa, L. and Palotás, K. and Udvardi, L. and Szunyogh, L.},
  title   = {Magnetic structure of monatomic \uppercase{F}e chains on \uppercase{R}e(0001): emergence of chiral multispin interactions},
  journal = {Phys. Rev. B},
  volume  = {99},
  pages   = {184430},
  year    = {2019},
  doi     = {10.1103/PhysRevB.99.184430}
}

@article{MoriyaYosida1953,
  author  = {Moriya, T. and Yosida, K.},
  title   = {On the origin of the anisotropy energy of \uppercase{C}u\uppercase{C}l$_2$·2\uppercase{H}$_2$\uppercase{O}},
  journal = {Prog. Theor. Phys.},
  volume  = {9},
  pages   = {663--675},
  year    = {1953},
  doi     = {10.1143/PTP.9.663}
}

@article{VanVleck1937,
  author  = {Van Vleck, J. H.},
  title   = {On the anisotropy of cubic ferromagnetic crystals},
  journal = {Phys. Rev.},
  volume  = {52},
  pages   = {1178--1198},
  year    = {1937},
  doi     = {10.1103/PhysRev.52.1178}
}

@article{Jackeli2009,
  author  = {Jackeli, G. and Khaliullin, G.},
  title   = {Mott insulators in the strong spin-orbit coupling limit: from \uppercase{H}eisenberg to a quantum compass and \uppercase{K}itaev models},
  journal = {Phys. Rev. Lett.},
  volume  = {102},
  pages   = {017205},
  year    = {2009},
  doi     = {10.1103/PhysRevLett.102.017205}
}

@article{Nussinov2015,
  author  = {Nussinov, Z. and van den Brink, J.},
  title   = {Compass models: theory and physical motivations},
  journal = {Rev. Mod. Phys.},
  volume  = {87},
  pages   = {1--59},
  year    = {2015},
  doi     = {10.1103/RevModPhys.87.1}
}

@article{Liechtenstein1987,
  author  = {Liechtenstein, A. I. and Katsnelson, M. I. and Antropov, V. P. and Gubanov, V. A.},
  title   = {Local spin density functional approach to the theory of exchange interactions in ferromagnetic metals and alloys},
  journal = {J. Magn. Magn. Mater.},
  volume  = {67},
  pages   = {65--74},
  year    = {1987},
  doi     = {10.1016/0304-8853(87)90076-1}
}

@article{Bruno2003,
  title = {Exchange \uppercase{I}nteraction \uppercase{P}arameters and \uppercase{A}diabatic \uppercase{S}pin-\uppercase{W}ave \uppercase{S}pectra of \uppercase{F}erromagnets: \uppercase{A} ``\uppercase{R}enormalized \uppercase{M}agnetic \uppercase{F}orce \uppercase{T}heorem''},
  author = {Bruno, P.},
  journal = {Phys. Rev. Lett.},
  volume = {90},
  issue = {8},
  pages = {087205},
  numpages = {4},
  year = {2003},
  month = {Feb},
  publisher = {American Physical Society},
  doi = {10.1103/PhysRevLett.90.087205},
  url = {https://link.aps.org/doi/10.1103/PhysRevLett.90.087205}
}

@article{Szilva2023,
  author  = {Szilva, A. and Kvashnin, Y. and Stepanov, E. A. and Nordstrom, L. and Eriksson, O. and Lichtenstein, A. I. and Katsnelson, M. I.},
  title   = {Quantitative theory of magnetic interactions in solids},
  journal = {Rev. Mod. Phys.},
  volume  = {95},
  pages   = {035004},
  year    = {2023},
  doi     = {10.1103/RevModPhys.95.035004}
}

@article{MartinezCarracedo2023,
  author  = {Martínez-Carracedo, G. and Oroszlány, L. and García-Fuente, A. and Nyári, B. and Udvardi, L. and Szunyogh, L. and Ferrer, J.},
  title   = {Relativistic magnetic interactions from nonorthogonal basis sets},
  journal = {Phys. Rev. B},
  volume  = {108},
  pages   = {214418},
  year    = {2023},
  doi     = {10.1103/PhysRevB.108.214418}
}

@article{Kurz2004,
  author  = {Kurz, Ph. and Förster, F. and Nordström, L. and Bihlmayer, G. and Blügel, S.},
  title   = {Ab initio treatment of noncollinear magnets with the full-potential linearized augmented plane wave method},
  journal = {Phys. Rev. B},
  volume  = {69},
  pages   = {024415},
  year    = {2004},
  doi     = {10.1103/PhysRevB.69.024415}
}

@article{Xiang2013,
  author  = {Xiang, H. and Lee, C. and Koo, H. and Gong, X. and Whangbo, M.},
  title   = {Magnetic properties and energy-mapping analysis},
  journal = {Dalton Trans.},
  volume  = {42},
  pages   = {823--853},
  year    = {2013},
  doi     = {10.1039/C2DT32210J}
}

@article{Sabani2020,
  author  = {Šabani, D. and Bacaksiz, C. and Milošević, M. V.},
  title   = {Ab initio methodology for magnetic exchange parameters: \uppercase{G}eneric four-state energy mapping onto a \uppercase{H}eisenberg spin \uppercase{H}amiltonian},
  journal = {Phys. Rev. B},
  volume  = {102},
  pages   = {014457},
  year    = {2020},
  doi     = {10.1103/PhysRevB.102.014457}
}

@article{Yang2015,
  author  = {Yang, H. and Thiaville, A. and Rohart, S. and Fert, A. and Chshiev, M.},
  title   = {Anatomy of \uppercase{D}zyaloshinskii–\uppercase{M}oriya interaction at \uppercase{C}o/\uppercase{P}t interfaces},
  journal = {Phys. Rev. Lett.},
  volume  = {115},
  pages   = {267210},
  year    = {2015},
  doi     = {10.1103/PhysRevLett.115.267210}
}

@article{Torelli2019,
  author  = {Torelli, D. and Thygesen, K. and Olsen, T.},
  title   = {High-throughput computational screening for 2\uppercase{D} ferromagnetic materials: the critical role of anisotropy and local correlations},
  journal = {2D Mater.},
  volume  = {6},
  pages   = {045018},
  year    = {2019},
  doi     = {10.1088/2053-1583/ab2f4f}
}

@article{Lounis2020,
  author  = {Lounis, S.},
  title   = {Multiple-scattering approach for multi-spin chiral magnetic interactions},
  journal = {New J. Phys.},
  volume  = {22},
  pages   = {103003},
  year    = {2020},
  doi     = {10.1088/1367-2630/abb3de}
}

@article{Xu2020,
  author  = {Xu, C. and Feng, J. and Kawamura, M. and Yamaji, Y. and Nahas, Y. and Prokhorenko, S. and Qi, Y. and Xiang, H. and Bellaiche, L.},
  title   = {Possible \uppercase{K}itaev quantum spin liquid state in 2\uppercase{D} materials with \uppercase{S} = 3/2},
  journal = {Phys. Rev. Lett.},
  volume  = {124},
  pages   = {087205},
  year    = {2020},
  doi     = {10.1103/PhysRevLett.124.087205}
}

@misc{SupplementaryInfo,
  title = {Supplementary \uppercase{I}nformation}
}

@book{Boyd2018,
  author  = {Boyd, S. and Vandenberghe, L.},
  title   = {Introduction to \uppercase{A}pplied \uppercase{L}inear \uppercase{A}lgebra},
  publisher = {Cambridge Univ. Press},
  year    = {2018}
}

@article{Zhang2021,
  author  = {Zhang, X. and Lu, Q. and Liu, W. and Niu, W. and Sun, J. and Cook, J. and Vaninger, M. and Miceli, P. and Singh, D. and Lian, S. and Chang, T. and He, X. and Du, J. and He, L. and Zhang, R. and Bian, G. and Xu, Y.},
  title   = {Room-temperature intrinsic ferromagnetism in epitaxial \uppercase{C}r\uppercase{T}e$_2$ ultrathin films},
  journal = {Nat. Commun.},
  volume  = {12},
  pages   = {2492},
  year    = {2021},
  doi     = {10.1038/s41467-021-22694-0}
}

@article{Meng2021,
  author  = {Meng, L. and Zhang, Z. and Xu, M. and Yang, S. and Si, K. and Liu, L. and Wang, X. and Jiang, H. and Li, B. and Qin, P. and Zhang, P. and Wang, J. and Liu, Z. and Tang, P. and Ye, Y. and Zhou, W. and Bao, L. and Gao, H. and Gong, Y.},
  title   = {Anomalous thickness dependence of Curie temperature in air-stable two-dimensional ferromagnetic \uppercase{C}r\uppercase{T}e$_2$ grown by chemical vapor deposition},
  journal = {Nat. Commun.},
  volume  = {12},
  pages   = {809},
  year    = {2021},
  doi     = {10.1038/10.1038/s41467-021-21072-z}
}

@article{Li2021,
  author  = {Li, B. et al.},
  title   = {Van der Waals epitaxial growth of air-stable \uppercase{C}r\uppercase{S}e$_2$ nanosheets},
  journal = {Nat. Mater.},
  volume  = {20},
  pages   = {818--825},
  year    = {2021},
  doi     = {10.1038/s41563-021-01038-6}
}

@article{Xian2022,
  author  = {Xian, J. and Wang, C. and Nie, J. and Li, R. and Han, M. and Lin, J. and Zhang, W. and Liu, Z. and Zhang, Z. and Miao, M. and Yi, Y. and Wu, S. and Chen, X. and Han, J. and Xia, Z. and Ji, W. and Fu, Y.},
  title   = {Spin mapping of intralayer antiferromagnetism in monolayer \uppercase{C}r\uppercase{T}e$_2$},
  journal = {Nat. Commun.},
  volume  = {13},
  pages   = {257},
  year    = {2022},
  doi     = {10.1038/s41467-021-27834-z}
}

@article{Wu2022,
  author  = {Wu, L. and Zhou, L. and Zhou, X. and Wang, C. and Ji, W.},
  title   = {In-plane epitaxy–strain-tuning magnetic coupling in \uppercase{C}r\uppercase{S}e$_2$ and \uppercase{C}r\uppercase{T}e$_2$ monolayers and bilayers},
  journal = {Phys. Rev. B},
  volume  = {106},
  pages   = {L081401},
  year    = {2022},
  doi     = {10.1103/PhysRevB.106.L081401}
}

@article{Kresse1996,
  author = {Kresse, G. and Furthm{\"u}ller, J.},
  title = {Efficient iterative schemes for ab initio total-energy calculations using a plane-wave basis set},
  journal = {Phys. Rev. B},
  volume = {54},
  pages = {11169--11186},
  year = {1996}
}

@article{Blochl1994,
  author = {Bl{\"o}chl, P. E.},
  title = {Projector augmented-wave method},
  journal = {Phys. Rev. B},
  volume = {50},
  pages = {17953--17979},
  year = {1994}
}

@article{Ma2015,
  author = {Ma, P. and Dudarev, S. L.},
  title = {Constrained density functional for noncollinear magnetism},
  journal = {Phys. Rev. B},
  volume = {91},
  pages = {054420},
  year = {2015}
}

@article{Xu2020a,
  author = {Xu, C. and Feng, J. and Prokhorenko, S. and Nahas, Y. and Xiang, H. and Bellaiche, L.},
  title = {Topological spin texture in \uppercase{J}anus monolayers of the chromium trihalides \uppercase{C}r(\uppercase{I},\uppercase{X})$_3$},
  journal = {Phys. Rev. B},
  volume = {101},
  pages = {060404},
  year = {2020}
}

@article{Xu2018,
  author = {Xu, C. and Feng, J. and Xiang, H. and Bellaiche, L.},
  title = {Interplay between \uppercase{K}itaev interaction and single-ion anisotropy in ferromagnetic \uppercase{C}r\uppercase{I}$_3$ and \uppercase{C}r\uppercase{G}e\uppercase{T}e$_3$ monolayers},
  journal = {npj Comput. Mater.},
  volume = {4},
  pages = {57},
  year = {2018},
  doi     = {10.1038/s41524-018-0115-6}
}

@article{Perdew1996,
  author = {Perdew, J. P. and Burke, K. and Ernzerhof, M.},
  title = {Generalized gradient approximation made simple},
  journal = {Phys. Rev. Lett.},
  volume = {77},
  pages = {3865--3868},
  year = {1996}
}

@article{Lu2023,
  author = {Lu, S. and Guo, D. and Cheng, Z. and Guo, Y. and Wang, C. and Deng, J. and Bai, Y. and Tian, C. and Zhou, L. and Shi, Y. and He, J. and Ji, W. and Zhang, C.},
  title = {Controllable dimensionality conversion between 1\uppercase{D} and 2\uppercase{D} \uppercase{C}r\uppercase{C}l$_3$ magnetic nanostructures},
  journal = {Nat. Commun.},
  volume = {14},
  pages = {2465},
  year = {2023}
}

@article{Huang2024,
  author = {Huang, B. and Chang, Y. and Lo, Y. and Fu, T.},
  title = {Behavior of iron deposition on the surface structure and electrical properties of \uppercase{C}r\uppercase{B}r$_3$ by scanning tunneling microscopy and spectroscopy},
  journal = {Thin Solid Films},
  volume = {800},
  pages = {140409},
  year = {2024}
}

@article{Huang2017,
  author = {Huang, B. and Clark, G. and Navarro-Moratalla, E. and Klein, D. and Cheng, R. and Seyler, K. and Zhong, D. and Schmidgall, E. and McGuire, M. and Cobden, D. and Yao, W. and Xiao, D. and Jarillo-Herrero, P. and Xu, X.},
  title = {Layer-dependent ferromagnetism in a van der Waals crystal down to the monolayer limit},
  journal = {Nature},
  volume = {546},
  pages = {270--273},
  year = {2017}
}

@article{McGuire2015,
  author = {McGuire, M. A. and Dixit, H. and Cooper, V. R. and Sales, B. C.},
  title = {Coupling of \uppercase{C}rystal \uppercase{S}tructure and \uppercase{M}agnetism in the \uppercase{L}ayered, \uppercase{F}erromagnetic \uppercase{I}nsulator \uppercase{C}r\uppercase{I}3},
  journal = {Chem. Mater.},
  volume = {27},
  pages = {612--620},
  year = {2015}
}

@article{Sun2021,
  author = {Sun, Y. and Yan, P. and Ning, J. and Zhang, X. and Zhao, Y. and Gao, Q. and Kanagaraj, M. and Zhang, K. and Li, J. and Lu, X. and Yan, Y. and Li, Y. and Xu, Y. and He, L.},
  title = {Ferromagnetism in two-dimensional \uppercase{C}r\uppercase{T}e$_2$ epitaxial films down to a few atomic layers},
  journal = {AIP Adv.},
  volume = {11},
  pages = {035138},
  year = {2021}
}

@article{Liu2021,
  author = {Liu, M. and Huang, Y. and Gou, J. and Liang, Q. and Chua, R. and Arramel, S. and Duan, S. and Zhang, L. and Cai, L. and Yu, X. and Zhong, D. and Zhang, W. and Wee, A. T. S.},
  title = {Diverse structures and magnetic properties in nonlayered monolayer chromium selenide},
  journal = {J. Phys. Chem. Lett.},
  volume = {12},
  pages = {7752--7760},
  year = {2021}
}

@article{Grimme2010,
  author = {Grimme, S. and Antony, J. and Ehrlich, S. and Krieg, H.},
  title = {A consistent and accurate ab initio parametrization of density functional dispersion correction (\uppercase{D}\uppercase{F}\uppercase{T}-\uppercase{D}) for the 94 elements \uppercase{H}--\uppercase{P}u},
  journal = {J. Chem. Phys.},
  volume = {132},
  pages = {154104},
  year = {2010}
}

@article{Hukushima1996,
  author = {Hukushima, K. and Nemoto, K.},
  title = {Exchange \uppercase{M}onte \uppercase{C}arlo method and application to spin glass simulations},
  journal = {J. Phys. Soc. Jpn.},
  volume = {65},
  pages = {1604--1608},
  year = {1996}
}

@article{Wang2026SpinGraph,
  author        = {Wang, Yifan and Deng, Boyang and Robertson, John and Zhao, Weisheng and Bl{\"u}gel, Stefan and Lu, Haichang},
  title         = {Automated \uppercase{G}eneration of \uppercase{C}ommensurate \uppercase{M}agnetic \uppercase{S}tructures based on \uppercase{S}pin \uppercase{S}pace \uppercase{G}roups and \uppercase{G}raph \uppercase{T}heory},
  year          = {2026},
  eprint        = {2609.09679},
  journal       = {arXiv},
  primaryClass  = {cond-mat.mtrl-sci},
  doi           = {10.48550/arXiv.2609.09679}
}

@article{Vosko:1980,
  title = {Influence of an improved local-spin-density correlation-energy functional on the cohesive energy of alkali metals},
  author = {Vosko, S. H. and Wilk, L.},
  journal = {Phys. Rev. B},
  volume = {22},
  issue = {8},
  pages = {3812--3815},
  numpages = {0},
  year = {1980},
  month = {Oct},
  publisher = {American Physical Society},
  doi = {10.1103/PhysRevB.22.3812},
  url = {https://link.aps.org/doi/10.1103/PhysRevB.22.3812}
}

\section*{Acknowledgements}
H.L.\ acknowledges fruitful discussions with Samir Lounis (Peter Grünberg Institute) on the microscopic origin of spin interactions, with Hongjun Xiang and Changsong Xu (Fudan University) on the computational algorithms, and with Gang Li (ShanghaiTech University) on group-theoretical analysis and acknowledges financial support from the National Key R\&D Programs of China (No.\ 2025YFA1411000 and 2024YFA1410200), the Beijing Natural Science Foundation (No.\ 2232055), and the National Natural Science Foundation of China (No.\ 12204027). S.B.\ and H.K.\ thank Juba Bouaziz and Nikolai Kiselev valuable discussions and for sharing their expertise on the calculation of Heisenberg interactions and topological indices. S.B.\ acknowledges financial support from the European Research Council (ERC) grant 856538 (project "3D MAGIC") and from the Deutsche Forschungsgemeinschaft (DFG, German Research Foundation) through SFB~1238 (Project C1).

\section*{Author Contributions}

Haichang Lu conceived the idea, developed \AMATIS, performed calculations, analyzed the data, wrote the paper, and supervised this project. Boyang Deng performed calculations and supported the data analysis. Hiroshi Katsumoto contributed to the theory of the quantum spin model. John Robertson and Weisheng Zhao provided constructive comments on the project. Stefan Bl\"{u}gel wrote the paper and co-supervised the project.

\section*{Code availability}

The source code and the necessary file have been deposited in the public GitHub (https://github.com/Haichang-Lu/AMATIS) and Zenodo (https://doi.org/10.5281/zenodo.16005430)  under the GNU General Public License version 3.0. 

\section*{Data availability}

All data supporting the findings of this study, including tensor components, magnetic configurations, and simulation outputs, are available within the article and its Supplementary Information. The tutorial, calculation data for demonstration, and the manual of \AMATIS~are available on the public Zenodo repository (https://doi.org/10.5281/zenodo.16005430) without any restrictions.

\section*{Competing financial interests}
The authors declare no competing interests.

\end{document}